\documentclass[10pt,journal]{IEEEtran}
\usepackage{amsmath,amsfonts}
\usepackage{algorithmic}
\usepackage{algorithm}
\usepackage{array}
\usepackage[caption=false,font=footnotesize,labelfont=sf,textfont=sf]{subfig}
\usepackage{textcomp}
\usepackage{stfloats}
\usepackage{url}
\usepackage{graphicx}
\usepackage{cite}
\usepackage{color} 
\usepackage{hyperref}
\hypersetup{pdfborder={0 0 0}}

\begin{document}

\title{Multi-Reference and Adaptive Nonlinear Transform Source-Channel Coding for Wireless Image Semantic Transmission}

\author{{Cheng Yuan,
	Yufei Jiang,~\IEEEmembership{Member,~IEEE,}
	and Xu Zhu~\IEEEmembership{Senior Member,~IEEE}
}

}



\maketitle

\begin{abstract}
We propose a multi-reference and adaptive nonlinear transform source-channel coding (MA-NTSCC) system for wireless image semantic transmission to improve rate-distortion (RD) performance by introducing multi-dimensional contexts into the entropy model of the state-of-the-art (SOTA) NTSCC system. 
Improvements in RD performance of the proposed MA-NTSCC system are particularly significant in high-resolution image transmission under low bandwidth constraints, compared with the existing NTSCC+ system. 
The proposed multi-reference entropy model leverages correlations within the latent representation in both spatial and channel dimensions.
In the spatial dimension, the latent representation is divided into anchors and non-anchors in a checkerboard pattern, where anchors serve as reference to estimate the mutual information between anchors and non-anchors.
In the channel dimension, the latent representation is partitioned into multiple groups, and features in previous groups are analyzed to estimate the mutual information between features in previous and current groups.
Taking mutual information into account, the entropy model provides an accurate estimation on the entropy, which enables efficient bandwidth allocation and enhances RD performance. 
Additionally, the proposed lightweight adaptation modules enable the proposed MA-NTSCC model to achieve transmission quality comparable to separately trained models across various channel conditions and bandwidth requirements.
In contrast, traditional NTSCC models provide signal-to-noise ratio (SNR)-distortion performance degrading with channel quality deviating from the fixed training SNR, and consume inflexible bandwidth to transmit an image.
The proposed adaptation modules leverage trainable splines which offer sufficient expressive power while maintaining low complexity.
Comprehensive experiments are conducted to verify the peak signal-to-noise ratio (PSNR) performance and adaptability of the proposed MA-NTSCC model superior to SOTA methods over both additive white Gaussian noise (AWGN) channel and Rayleigh fading channel.
\end{abstract}

\section{Introduction}
In traditional communication systems, the encoder consists of three separate modules: source coding, channel coding and modulation.
Source coding compresses source data while retaining information necessary for reconstruction at a certain fidelity.
Channel coding introduces structured redundancy to combat channel impairments. 
Modulation converts the bit stream into transmitted symbols.
The separate design simplifies system development and deployment, but consumes excessive bandwidth, and thus is transmission-inefficient \cite{survey1}.
Moreover, modern source coding and channel coding techniques are highly sophisticated, leading to high communication latency.
For instance, the encoder of versatile video coding (VVC) \cite{VVC} evaluates multiple choices of transform types and unit sizes, and the decoder of the low-density parity check (LDPC) code leverages high-complexity iterations \cite{LDPC}, both incurring extensive computation.
Additionally, this separate coding paradigm suffers from the cliff effect \cite{survey2}: When channel quality falls below a certain threshold, the channel decoder cannot recover the bit stream correctly, causing the source decoder to fail in reconstructing meaningful data.
To ensure successful transmission over a volatile wireless channel, hybrid automatic repeat request (HARQ) is employed, which further increases latency.
These limitations hinder the performance of latency-sensitive and reliability-critical tasks, such as autonomous driving, drone surveillance and extended reality (XR).

Joint source-channel coding (JSCC), as an emerging paradigm in semantic and task-oriented communications, overcomes limitations in communication latency and bandwidth consumption \cite{survey1, survey2, survey3, survey4, survey5}.
Deep learning (DL)-based JSCC leverages deep neural networks (DNNs) to directly transform source data into transmitted symbols, allowing graceful degradation in performance over ill-conditioned channels, immune from the cliff effect \cite{DJSCC}.
JSCC simplifies the implementation of the communication system, and reduces system latency.
Additionally, the JSCC system improves rate-distortion (RD) performance by jointly optimizing data compression and channel coding, thereby conserving bandwidth.
The JSCC system has been shown effective in various communication scenarios, such as feedback channel \cite{DJSCC-f}, progressive transmission \cite{DJSCC-l}, finite alphabet \cite{DJSCC-Q} and multiple access \cite{DeepMA}.
It is also compatible with widely-employed orthogonal frequency division multiplexing (OFDM) \cite{OFDM} and multiple-input multiple-output (MIMO) \cite{MIMO, CSI}.
The performance of the JSCC system can be further enhanced by incorporating other image processing networks.
In \cite{GAN}, a generative adversarial network (GAN) is integrated into the JSCC architecture to improve perceptual quality of received images in low signal-to-noise ratio (SNR) and low bandwidth regimes.
In \cite{CDDM}, a diffusion model is employed at receiver to mitigate channel noise, thus enhancing the quality of reconstructed images.

The aforementioned JSCC systems encode each patch of an image at a uniform rate, disregarding the significant variation in semantic complexity across patches.
Consequently, the performances of these JSCC systems are inferior to those of state-of-the-art (SOTA) separate coding scheme that combines VVC with LDPC.
To address this problem, two variable-length coding schemes have been proposed for the JSCC architecture: a data-driven approach \cite{XR} and a model-driven approach \cite{NTSCC}.
In \cite{XR}, a rate allocation network is utilized at transmitter to generate a binary mask for rate control.
The nonlinear transform source-channel coding (NTSCC) \cite{NTSCC} utilizes a SOTA neural image compression model, referred to as nonlinear transform coding (NTC) \cite{NTC_survey, Balle2018}, to transform the source image into a latent representation whose entropy is estimated by modeling its probability distribution.
Subsequently, rate allocation is conducted based on the estimated entropy, and an encoder network converts the latent representation into transmitted symbols of variable lengths.
Inspired by advancements in NTC models \cite{Cheng2020, checkerboard}, the improved NTSCC model \cite{NTSCC+} leverages checkerboard context in both the NTC model and the JSCC network to achieve RD performance surpassing SOTA separate coding methods.
However, implementing checkerboard context in the JSCC network restricts information exchange when encoding and decoding the first half of symbols.
Consequently, the correlations between latent features are not fully exploited, and the performance gains almost vanish in high-resolution image transmission.

Furthermore, mainstream JSCC models are trained for specific code rates and channel qualities, causing performance degradation when channel conditions change during testing.
Thus, multiple sets of weights must be trained under different code rates and SNRs, and  synchronized weight switching at both transmitter and receiver is required to adapt to dynamic channel conditions and bandwidth requirements in wireless communications.
To maintain model performance across varying environments, adaptation modules have been proposed in the literature \cite{ADJSCC,DJSCC-V,NTSCC+,ModNet}.
In \cite{ADJSCC}, multiple attention feature (AF) modules based on fully connected (FC) networks are employed in the JSCC model, assigning channel-wise weights to hidden features based on their averages and SNR.
In \cite{ModNet}, two FC networks are leveraged to modulate encoded symbols and received symbols according to SNR, respectively.
However, the FC networks introduce an excessive number of parameters and high computational complexity \cite{ADJSCC, ModNet}.
In \cite{DJSCC-V}, a spatially invariant binary mask is applied to the encoded symbols according to the selected code rate, enabling multiple rate options for a single model.
However, this method is incompatible with the NTSCC system, where bandwidth is allocated based on the estimated entropy.
In \cite{NTSCC+}, channel-wise weighting vectors are employed on latent features and encoded symbols in the NTSCC system.
However, the expressive power of weighting vectors is insufficient to effectively adapt a single model to dynamic environments, resulting in performance losses compared to separately trained models, particularly for high-resolution images.

Motivated by the open issues above, we propose a multi-reference and adaptive NTSCC (MA-NTSCC) system to improve RD performance for wireless image semantic transmission by introducing multi-dimensional contexts into the entropy model of the existing NTSCC system \cite{NTSCC}.
Also, we design lightweight adaptation modules to guarantee model performance of the proposed MA-NTSCC system across various channel conditions and bandwidth requirements.
Main contributions of this work are summarized as follows:
\begin{itemize}
    \item We propose a multi-reference NTSCC (MR-NTSCC) system, which leverages correlations within the latent representation in both spatial and channel dimensions.
    In the spatial dimension, the latent representation is divided into anchors and non-anchors in a checkerboard pattern, where anchors serve as references to estimate the mutual information between anchors and non-anchors.
    In the channel dimension, the latent representation is partitioned into multiple groups, and features in previous groups are analyzed to estimate the mutual information between features in previous and current groups.
    Taking mutual information into account, the proposed model provides an accurate estimation on the entropy, which enables efficient bandwidth allocation and enhances RD performance. 
    Improvements in RD performance are particularly significant in high-resolution image transmission under low bandwidth constraints, compared with the NTSCC+ system \cite{NTSCC+}.

    \item Based on the proposed MR-NTSCC system, we design lightweight adaptation modules to achieve transmission quality comparable to separately trained models across different SNRs and bandwidth ratios.
    In contrast, traditional NTSCC models \cite{NTSCC} provide SNR-distortion performance degrading with channel quality deviating from the fixed training SNR, and consume inflexible bandwidth to transmit an image.
    The proposed adaptation modules leverage trainable splines which offer sufficient expressive power while maintaining low complexity.
    Three types of adaptation modules are introduced into the proposed MA-NTSCC model to enhance adaptability.
    First, the rate adaptation (RA) module modifies hidden features in the NTC model to meet bandwidth requirements.
    Second, the adaptive embedding (AE) module provides encoder and decoder networks with information on channel qualities and code rates.
    Third, the adaptive transform (AT) module adjusts hidden features in the encoder and decoder networks to accommodate dynamic SNRs and bandwidth ratios.

    \item Comprehensive experiments are conducted to verify the performance and adaptability of the proposed MA-NTSCC model.
    The proposed MA-NTSCC model achieves up to 50\% bandwidth savings at identical transmission quality, or up to 6 dB improvement in peak signal-to-noise ratio (PSNR) with equal SNR and bandwidth consumption, compared with the compatible NTSCC+ model \cite{NTSCC+}, over both additive white Gaussian noise (AWGN) channel and Rayleigh fading channel.
    Visual comparison is conducted to demonstrate perceptual quality, and trained spline functions are presented and analyzed.
    Ablation study validates the bandwidth savings provided by the proposed multi-reference entropy model and the adaptability of the proposed adaptation modules.
    Complexity analysis of the proposed MA-NTSCC model shows a 30\% reduction in model size and a 34\% decrease in runtime, compared with the compatible NTSCC+ model \cite{NTSCC+}.
\end{itemize}


\section{System Model}
\begin{figure}[!t]
    \centering
    \includegraphics[width=1\linewidth]{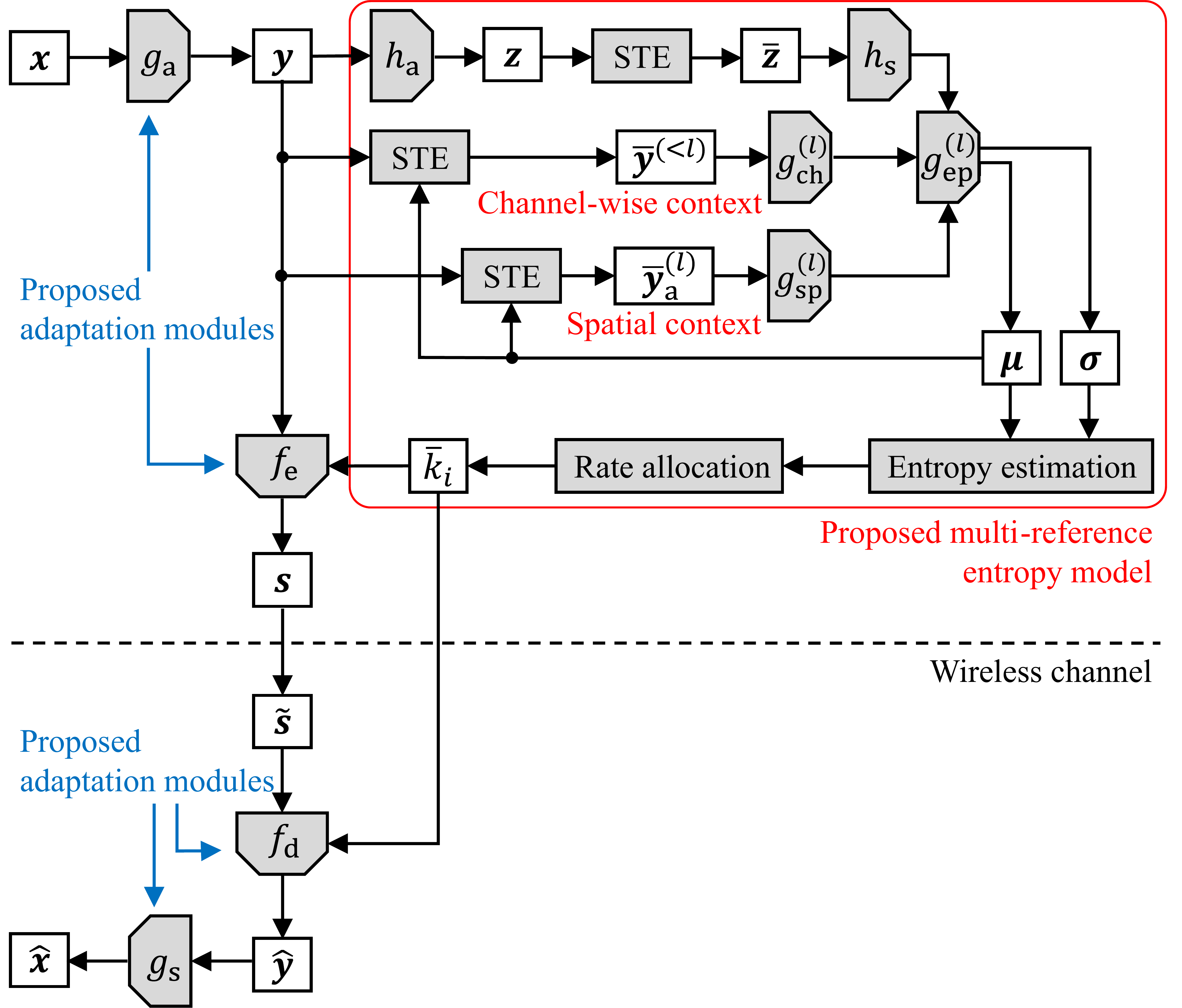}
    \caption{Block diagram of the proposed MA-NTSCC system. 
    (We propose to leverage correlated contexts in both spatial and channel dimensions to accurately estimate the entropy of latent features $\boldsymbol{y}$, thereby improving the efficiency of rate allocation and enhancing the RD performance.
    We also design three types of lightweight adaptation modules to enable a single MA-NTSCC model to achieve transmission quality comparable to separately trained models across different SNRs and bandwidth ratios.)}
    \label{fig_sys}
\end{figure}
As illustrated in Fig. \ref{fig_sys}, we consider a NTSCC system for wireless image semantic transmission.
Given an image represented by a vector of pixel-wise RGB intensities $\boldsymbol{x} \in \mathbb{R}^m$ with $m$ denoting the length of the image vector, the image $\boldsymbol{x}$ is reshaped into a size of $h \times w \times 3$, where $h$ and $w$ represent the height of and the width of the image, respectively. 
If the height of or the width of the input image is not divisible by 16, padding is applied to meet this requirement, which is removed from the decoded image.
A DNN-based nonlinear analysis transform $g_{\rm{a}}$ is employed to extract the semantic information in the image. 
A latent representation $\boldsymbol{y} \in \mathbb{R}^{\frac{h}{16} \times \frac{w}{16} \times c}$ is obtained, with $c$ denoting the channel dimension, due to down-sampling operation performed by the analysis transform $g_{\rm{a}}$.
The latent representation $\boldsymbol{y}$ can be partitioned into $L$ groups in the channel dimension as: 
\begin{equation}
    \boldsymbol{y} = \{ \boldsymbol{y}^{(1)}, \boldsymbol{y}^{(2)}, \ldots, \boldsymbol{y}^{(L)} \} ,
    \label{eq_K_groups}
\end{equation}
where $\boldsymbol{y}^{(l)} \in \mathbb{R}^{\frac{h}{16} \times \frac{w}{16} \times c_l}$ represents the $l$-th group of latent features, with $c_l$ denoting the channel dimension of the $l$-th group.
The latent representation can also be viewed as a collection of vectors, each with a length of $c$ as $\boldsymbol{y}_i \in \mathbb{R}^c$, $i = 1, 2, \ldots, \frac{hw}{256}$.
Each vector $\boldsymbol{y}_i$ is a compact representation of semantic information in a $16 \times 16$ patch of the input image.
Moreover, due to the global receptive field of convolutional neural networks (CNNs) in the analysis transform $g_{\rm{a}}$, $\boldsymbol{y}_i$ also encapsulates semantic information at a large scale.

To facilitate bandwidth allocation, the entropy of $\boldsymbol{y}$ is estimated by a learned entropy model.
The probability distribution of $\boldsymbol{y}$ is modeled as Gaussian with its mean and variance conditioned on a hyperprior $\boldsymbol{z}$ \cite{Balle2018}, written as:
\begin{equation}
    p_{{\boldsymbol{y}} | {\boldsymbol{z}}} ({\boldsymbol{y}} | {\boldsymbol{z}}) = 
    \mathcal{N}(\boldsymbol{\mu}, \boldsymbol{\sigma}^2)({\boldsymbol{y}}) ,
    \label{eq_prob_y}
\end{equation}
where $\mathcal{N}(\boldsymbol{\mu}, \boldsymbol{\sigma}^2)$ denotes factorized Gaussian distribution with a mean vector of $\boldsymbol{\mu}$ and a variance vector of $\boldsymbol{\sigma}^2$.
The hyperprior $\boldsymbol{z}$ is obtained from $\boldsymbol{y}$ by a hyperprior analysis transform $h_{\rm{a}}$, and is processed by a hyperprior synthesis transform $h_{\rm s}$.
The means and the standard deviations of $\boldsymbol{y}$ are estimated from the hyperprior $\boldsymbol{z}$ by a subnet $g_{\rm ep}$, written as:
\begin{equation}
    (\boldsymbol{\mu}, \boldsymbol{\sigma}) = g_{\rm ep}( h_{\rm{s}}({\boldsymbol{z}}) ) .
    \label{eq_h_s}
\end{equation}
During model training, it is beneficial to employ straight-through estimator (STE) on the latent representation $\boldsymbol{y}$ and the hyperprior $\boldsymbol{z}$ \cite{Minnen2020}.
Specifically, the latent representation $\boldsymbol{y}$ is quantized after obtaining the mean vector $\boldsymbol{\mu}$ as:
\begin{equation}
    \bar{\boldsymbol{y}} = \rm{STE}(\boldsymbol{y} - \boldsymbol{\mu}) + \boldsymbol{\mu} ,
    \label{eq_bar_y}
\end{equation}
where ${\rm STE}(x) = {\rm sg}({\rm round}(x) - x) + x$ is the STE operation \cite{TOFC}, ${\rm sg}(\cdot)$ is the stop-gradient operator, and ${\rm round}(\cdot)$ denotes rounding to the nearest integer.
The hyperprior $\boldsymbol{z}$ is quantized similarly as:
\begin{equation}
    \bar{\boldsymbol{z}} = \rm{STE}(\boldsymbol{z} - \boldsymbol{\mu}_{\rm{1/2}}) + \boldsymbol{\mu}_{\rm{1/2}} ,
    \label{eq_bar_z}
\end{equation}
where $\boldsymbol{\mu}_{\rm{1/2}}$ denotes the median vector of $\boldsymbol{z}$, obtained from a learned probability model \cite{Balle2018}.
After training finishes, whether STE is employed during testing has negligible impact on the RD performance, thus we omit STE during testing.

The NTSCC system utilizes a trainable DNN to achieve JSCC, where ${\boldsymbol{y}}$ is directly mapped into symbols for wireless transmissions $\boldsymbol{s} \in \mathbb{C}^{k}$, with $k$ denoting the number of complex symbols.
The channel bandwidth ratio (CBR) is defined as $R = k/m$ to evaluate the average coding rate of an image \cite{NTSCC}.
The entropy of $\boldsymbol{y}$ is estimated from the probability model of the quantized ${\boldsymbol{y}}$ as $- \mathrm{log} \, p_{{\boldsymbol{y}} | {\boldsymbol{z}}} ({\boldsymbol{y}} | {\boldsymbol{z}})$, and is then used for bandwidth allocation in the encoder network $f_{\rm{e}}$.
The bandwidth allocated to the $i$-th latent vector is determined by its estimated entropy as \cite{NTSCC}:
\begin{equation}
    k_i = \eta \sum_{j=1}^{c} -\log p_{{y}_{i,j} | {\boldsymbol{z}}} (y_{i,j} | \boldsymbol{z}) ,
    \label{eq_k_i}
\end{equation}
where $\eta$ is a bandwidth coefficient representing the scaling relation between the allocated bandwidth and the estimated entropy of the latent representation, and $y_{i,j}$ denotes the $j$-th $(j = 1, 2, \ldots, c)$ element of $\boldsymbol{y}_i$.
The symbol length $k_i$ is quantized to a discrete set of possible symbol lengths $\mathcal{K} = \{ v_1, v_2, \ldots, v_K \}$ as $\bar{k}_i$.
The complex symbols $\boldsymbol{s}$ generated by the encoder $f_{\rm{e}}$ can be viewed as a collection of vectors $\boldsymbol{s}_i$, each with a length of $\bar{k}_i$.
The length of each symbol vector $\bar{k}_i$ is transmitted to the receiver as side information.
The received symbols $\tilde{\boldsymbol{s}}$ over fading channels are expressed as:
\begin{equation}
    \tilde{\boldsymbol{s}} = \boldsymbol{h} \odot \boldsymbol{s} + \boldsymbol{n} ,
    \label{eq_channel}
\end{equation}
where $\boldsymbol{h}$ is the channel state information (CSI) vector, $\odot$ denotes element-wise product, and $\boldsymbol{n}$ denotes complex AWGN.
At receiver, the latent representation is recovered from the received symbols by a DNN-based decoder $f_{\rm{d}}$ as $\hat{\boldsymbol{y}} = f_{\rm{d}}(\tilde{\boldsymbol{s}})$.
Finally, the synthesis transform is employed to reconstruct the original image as $\hat{\boldsymbol{x}} = g_{\rm{s}}(\hat{\boldsymbol{y}})$.

To achieve a trade-off between data rate and transmission quality, we employ RD optimization for model training.
The loss function is given by:
\begin{equation}
    \mathcal{L}_{\rm NTSCC} = -\eta \sum_{i} \log p_{\boldsymbol{y}_i | \boldsymbol{z}}(\boldsymbol{y}_i | \boldsymbol{z}) + \lambda d(\boldsymbol{x}, \hat{\boldsymbol{x}}) ,
    \label{eq_loss_NTSCC}
\end{equation}
where $\lambda$ is a Lagrange multiplier that controls the trade-off between rate and distortion, and $d(\cdot, \cdot)$ denotes the distortion function.

\section{Multi-Reference NTSCC}
We propose a MR-NTSCC system, where correlated contexts in both the spatial and channel dimensions are utilized as references to accurately estimate the entropy of the latent representation $\boldsymbol{y}$, thus enhancing the RD performance.
This is different from conventional NTSCC systems, which estimate the entropy of latent features from only the factorized hyperprior $\boldsymbol{z}$, disregarding the correlations in two dimensions.
As a result, the estimated entropy does not take the mutual information within latent features into account, and exceeds the actual value, which impedes efficient rate allocation and degrades the RD performance.

The feasibility of our method stems from the semantic characteristics of images.
Images show strong correlations in local regions, and such correlations partially persist after the convolution operations \cite{checkerboard}.
Consequently, the latent representation generated by CNNs in the analysis transform $g_{\rm a}$ is still correlated.
We propose to leverage DNNs to extract semantic information from correlated contexts in two dimensions: spatial dimension and channel dimension \cite{ELIC}.
This information assists the entropy model in analyzing the mutual information between latent features, thereby obtaining an accurate estimation of the entropy.
When estimating the entropy of the $i$-th latent vector $\boldsymbol{y}_i$, the spatial correlations allow adjacent latent vectors $\boldsymbol{y}_{n}$ $(n \neq i)$ to serve as references for the entropy model, referred to as spatial context.
Additionally, when estimating the entropy of the $l$-th group of the latent representation $\boldsymbol{y}^{(l)}$, the entropy model can utilize features in previous groups $\boldsymbol{y}^{(<l)}$ as channel-wise context.
With more precise estimates of the entropy, the ensuing encoder can allocate bandwidth more efficiently, which boosts RD performance.
Improvements in RD performance are particularly significant when the image resolution is high, and when channel bandwidth available is limited.

In the following subsections, we first introduce DNNs designed to extract semantic information from contexts in spatial and channel dimensions. 
Subsequently, we discuss how the entropy model integrates these semantic information with the hyperprior $\boldsymbol{z}$.

\subsection{Spatial Context}
\begin{figure}[!t]
    \centering
    \includegraphics[width=1\linewidth]{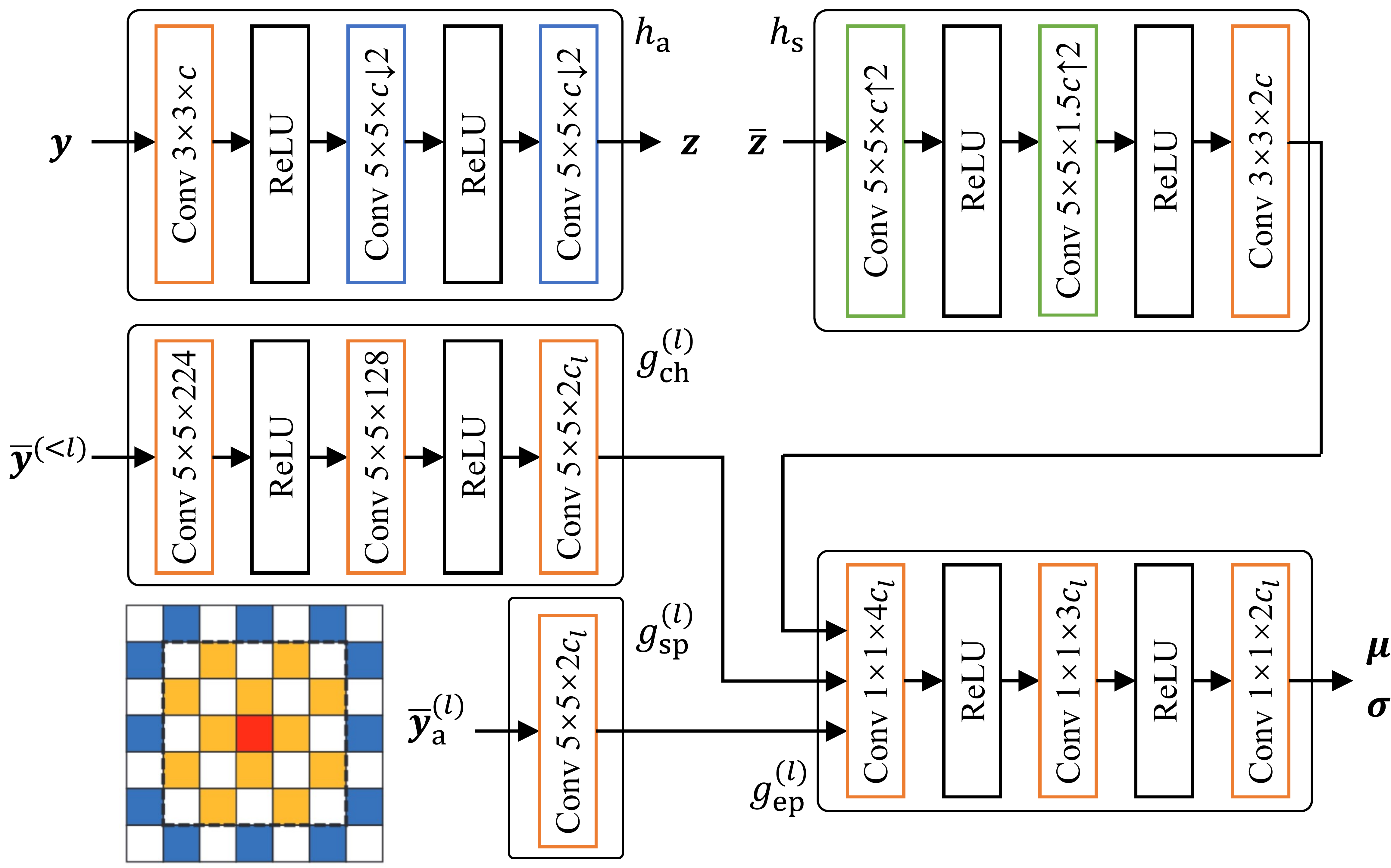}
    \caption{Network architecture of the proposed multi-reference entropy model, which leverages correlated contexts in both spatial and channel dimensions to accurately estimate the entropy of latent features $\boldsymbol{y}$. The spatial context is extracted from neighboring anchors $\bar{\boldsymbol{y}}_{\rm a}^{(l)}$ (colored in orange) by a subnet $g_{\rm sp}^{(l)}$ to facilitate the entropy estimation of non-anchors (colored in red). The channel-wise context is extracted from features in previous groups $\bar{\boldsymbol{y}}^{(<l)}$ by another subnet $g_{\rm ch}^{(l)}$ to assist the entropy estimation of the current group.
    Parameters of convolutional layer are denoted as kernel height $\times$ kernel width $\times$ the number of filters, $\uparrow$ and $\downarrow$ denote up-sampling and down-sampling, respectively, followed by the stride.}
    \label{fig_entropy_model}
\end{figure}
As illustrated in Fig. \ref{fig_entropy_model}, the latent representation $\boldsymbol{y}$ is divided into two parts in the spatial dimension: one half of the features (colored in orange and blue in Fig. \ref{fig_entropy_model}) are referred to as anchors, and the other half of the features (colored in white and red in Fig. \ref{fig_entropy_model}) are referred to as non-anchors. 
We employ two-pass computation with checkerboard pattern to obtain spatial context to facilitate parallel computation of graphics processing units (GPUs).
In the first pass, the statistics of anchors are obtained only from the hyperprior as:
\begin{equation}
    (\boldsymbol{\mu}_{\rm a}, \boldsymbol{\sigma}_{\rm a}) = g_{\rm ep,a}(h_{\rm s}(\bar{\boldsymbol{z}})) ,
    \label{eq_g_sp,a}
\end{equation}
As mentioned in Section II, the anchors are quantized by STE as in (\ref{eq_bar_y}) during training.
In the second pass, the semantic information already encoded in anchors is analyzed by a subnet $g_{\rm sp}$.
Subsequently, the statistics of non-anchors are calculated from both the hyperprior and the semantic information in anchors as:
\begin{equation}
    (\boldsymbol{\mu}_{\rm na}, \boldsymbol{\sigma}_{\rm na}) = g_{\rm ep,na}(h_{\rm s}(\bar{\boldsymbol{z}}), g_{\rm sp}(\bar{\boldsymbol{y}}_{\rm a})) ,
    \label{eq_g_sp}
\end{equation}
where $\boldsymbol{\mu}_{\rm na}$ and $\boldsymbol{\sigma}_{\rm na}$ denote the means and the standard deviations of non-anchors, and $\bar{\boldsymbol{y}}_{\rm a}$ denotes the quantized latent representation with the positions of anchors filled with actual values and the positions of non-anchors filled with zeros.
As illustrated in Fig. \ref{fig_entropy_model}, we choose convolution operations as the foundation of the subnet $g_{\rm sp}$ instead of global attention, based on the semantic characteristics of images.
As correlations between different patches of images rapidly diminish with the increasing distance \cite{checkerboard}, the mutual information between anchors and non-anchors can be accurately analyzed from only the anchors in the vicinity of each non-anchor.
When encoding the non-anchor latent vector colored in red in Fig. \ref{fig_entropy_model}, only the neighboring anchors colored in orange are analyzed by the subnet $g_{\rm sp}$, which significantly reduces the computational complexity and the number of parameters, compared with global attention.

From the perspective of information theory, the actual entropy of the latent representation, considering correlations in the spatial dimension, is given by:
\begin{equation}
    H(\boldsymbol{y}) = H(\boldsymbol{y}_{\rm a}) + H(\boldsymbol{y}_{\rm na}) - I(\boldsymbol{y}_{\rm a}, \boldsymbol{y}_{\rm na}) ,
    \label{eq_I_sp}
\end{equation}
where $H(\cdot)$ denotes the entropy, and $I(a, b)$ denotes the mutual information between $a$ and $b$.
The entropy model cannot estimate the mutual information from factorized hyperprior $\boldsymbol{z}$, and the estimated entropy is close to $H(\boldsymbol{y}_{\rm a}) + H(\boldsymbol{y}_{\rm na})$, higher than its actual value as in (\ref{eq_I_sp}).
However, provided with the semantic information in anchors, the entropy model can estimate the mutual information between anchors and non-anchors $I(\boldsymbol{y}_{\rm a}, \boldsymbol{y}_{\rm na})$ when estimating the entropy of non-anchors, thereby achieving a more accurate estimation of the joint entropy.

\subsection{Channel-wise Context}
The latent representation $\boldsymbol{y}$ is partitioned into $L$ groups along the channel dimension as in (\ref{eq_K_groups}), and each group is encoded sequentially.
Starting from the second group, the features in previously encoded groups serve as channel-wise context.
When estimating the entropy of the $l$-th ($l = 2, 3, \ldots, L$) group of latent features, the semantic information already encoded in previous groups of features is analyzed by a subnet $g_{\rm ch}^{(l)}$.
As the sizes of the input channel-wise context and the output statistics vary with group index, each group employs distinct subnets $g_{\rm ch}^{(l)}$ and $g_{\rm ep}^{(l)}$.
Thus, the statistics of each group of latent features are calculated as:
\begin{equation}
    (\boldsymbol{\mu}^{(l)}, \boldsymbol{\sigma}^{(l)}) = \begin{cases}
        { g_{\rm ep}^{(l)}(h_{\rm s}(\bar{\boldsymbol{z}})) ,} & {l = 1} \\
        { g_{\rm ep}^{(l)}(h_{\rm s}(\bar{\boldsymbol{z}}), g_{\rm ch}^{(l)}(\bar{\boldsymbol{y}}^{(<l)})) ,} & {l = 2, 3, \ldots, L}
    \end{cases}
    ,
    \label{eq_g_ch}
\end{equation}
where $\boldsymbol{\mu}^{(l)}$ and $\boldsymbol{\sigma}^{(l)}$ denote the means and standard deviations of latent features in the $l$-th group.
As illustrated in Fig. \ref{fig_entropy_model}, we choose CNNs as the foundation of the subnet $g_{\rm ch}$ instead of transformer blocks.
The receptive field of CNNs with a relatively large kernel size of $5 \times 5$ is sufficient to capture the semantic information encoded in previous groups.
Additionally, the computational complexity of convolution is lower than that of multi-head self attention (MSA) utilized in transformer blocks, as convolution operations are limited to the kernel size, while attention is computed between any two patches across the entire input.
To further reduce complexity, we progressively decrease the number of channels layer by layer, which is particularly effective in the later groups due to the large number of channels of features in previous groups.

Considering correlations in the channel dimension, the actual entropy of the latent representation is given by:
\begin{equation}
    H(\boldsymbol{y}) = \sum_{l=1}^{L} H(\boldsymbol{y}^{(l)}) - \sum_{l=2}^{L} I(\boldsymbol{y}^{(l)}, \boldsymbol{y}^{(<l)}) .
    \label{eq_I_ch}
\end{equation}
The entropy model can only estimate the individual entropy of features in each group $H(\boldsymbol{y}^{(l)})$ from factorized hyperprior $\boldsymbol{z}$.
However, the semantic information extracted from previously encoded groups assists the entropy model in estimating the mutual information between the current group of latent features and previously encoded groups $I(\boldsymbol{y}^{(l)}, \boldsymbol{y}^{(<l)})$, thereby achieving a more accurate estimation of the joint entropy.

\subsection{Multi-Reference Entropy Model}
The proposed MR-NTSCC system employs a multi-reference entropy model, which combines spatial and channel-wise contexts with factorized hyperprior to accurately estimate the joint entropy of the latent representation.
The statistics of the latent representation are calculated group by group in the channel dimension, and the latent features within each group are divided into anchors and non-anchors according to the checkerboard pattern.
Starting from the second group, the semantic information encoded in features in previous groups is extracted by a subnet $g_{\rm ch}$ and utilized as channel-wise context when calculating the statistics of both anchors and non-anchors.
For each group, the statistics of latent features are calculated in two passes.
In the first pass, the statistics of anchors are obtained from the hyperprior and the channel-wise context.
In the second pass, the semantic information encoded in anchors is analyzed by a subnet $g_{\rm sp}$ and employed in the calculation of the statistics of non-anchors.
The calculations of the statistics of anchors and non-anchors in the $l$-th group are respectively given by:
\begin{equation}
    (\boldsymbol{\mu}_{\rm a}^{(l)}, \boldsymbol{\sigma}_{\rm a}^{(l)}) = \begin{cases}
        { g_{\rm ep,a}^{(l)}(h_{\rm s}(\bar{\boldsymbol{z}})) ,} & {l = 1} \\
        { g_{\rm ep,a}^{(l)}(h_{\rm s}(\bar{\boldsymbol{z}}), g_{\rm ch}^{(l)}(\bar{\boldsymbol{y}}^{(<l)})) ,} & {l > 1}
    \end{cases}
    ,
    \label{eq_h_s,a}
\end{equation}
\begin{equation}
    (\boldsymbol{\mu}_{\rm na}^{(l)}, \boldsymbol{\sigma}_{\rm na}^{(l)}) = \begin{cases}
    {g_{\rm ep,na}^{(l)}(h_{\rm s}(\bar{\boldsymbol{z}}), g_{\rm sp}^{(l)}(\bar{\boldsymbol{y}}_{\rm a})) ,} & {l = 1} \\
    {g_{\rm ep,na}^{(l)}(h_{\rm s}(\bar{\boldsymbol{z}}), g_{\rm ch}^{(l)}(\bar{\boldsymbol{y}}^{(<l)}), g_{\rm sp}^{(l)}(\bar{\boldsymbol{y}}_{\rm a})) ,} & {l > 1}
    \end{cases}
    .
    \label{eq_h_s,na}
\end{equation}
As illustrated in Fig. \ref{fig_entropy_model}, we employ $1 \times 1$ CNNs as the foundation of the subnet $g_{\rm ep}$ to reduce complexity.
The model's performance is not compromised by the small kernel size, as the necessary semantic information in the vicinity is extracted by subnets $g_{\rm sp}$ and $g_{\rm ch}$, which utilize large kernel size.
The numbers of channels of hidden features in $g_{\rm ep}$ are linearly interpolated between the numbers of channels of the input and the output.

Considering correlations in both spatial and channel dimensions, the actual entropy of the latent representation is given by:
\begin{align}
    H(\boldsymbol{y}) = & \sum_{l=1}^{L} (H(\boldsymbol{y}_{\rm a}^{(l)}) + H(\boldsymbol{y}_{\rm na}^{(l)})) \nonumber \\
    & - \sum_{l=2}^{L} (I(\boldsymbol{y}_{\rm a}^{(l)}, \boldsymbol{y}_{\rm na}^{(l)}) + I(\boldsymbol{y}^{(l)}, \boldsymbol{y}^{(<l)})) .
    \label{eq_I_both}
\end{align}
The entropy model can only estimate the individual entropy of anchors and non-anchors in each group, namely the first item in (\ref{eq_I_both}), from factorized hyperprior $\boldsymbol{z}$.
However, the proposed MR-NTSCC system utilizes correlated contexts in both spatial and channel dimensions to thoroughly estimate the mutual information, consisting of two components: the mutual information between anchors and non-anchors in the current group $I(\boldsymbol{y}_{\rm a}^{(l)}, \boldsymbol{y}_{\rm na}^{(l)})$, and the mutual information between the current group of latent features and previous groups $I(\boldsymbol{y}^{(l)}, \boldsymbol{y}^{(<l)})$.
Thus, the proposed MR-NTSCC system accurately estimates the actual entropy of latent features and enhances the efficiency of rate allocation during feature encoding.

\section{Adaptive NTSCC}
We propose lightweight adaptation modules to enable a single MA-NTSCC model to deliver competitive performance across various SNRs and bandwidth ratios, comparable to that of separately trained models.
In contrast, traditional NTSCC models \cite{NTSCC} provide SNR-distortion performance degrading with channel quality deviating from the fixed training SNR, and consume inflexible bandwidth to transmit an image.
Thus, multiple sets of weights must be trained under different SNRs and code rates, and  synchronized weight switching at both transmitter and receiver is required to adapt to dynamic channel conditions and bandwidth requirements in wireless communications.
On the contrary, the proposed MA-NTSCC model maintains performance in dynamic environments, and is highly practical for wireless image transmission with volatile channel conditions and bandwidth restrictions.

The proposed adaptation modules leverage trainable splines, \textit{i.e.}, piecewise linear functions, instead of commonly used FC networks \cite{ADJSCC, DJSCC-V, ModNet}.
A trainable spline designates $N$ knots, each with a fixed input value $x_i$ in the ascending order and a learnable output value $f(x_i)$, $i = 1, 2, \ldots, N$.
Given an input $x$, the output $f(x)$ is computed by linear interpolation between the output values of the two closest knots as:
\begin{equation}
    f(x) = \begin{cases}
        {f(x_1) ,} & {x < x_1} \\
        {\frac{f(x_{i+1} - f(x_i)}{x_{i+1} - x_i}(x-x_i) ,} & {x_i < x < x_{i+1}} \\
        {f(x_N) ,} & {x > x_N}
    \end{cases}
    .
    \label{eq_spline}
\end{equation}
During the training process, the output values of these knots are adjusted to optimize the performance of the model across different SNRs and bandwidth ratios, allowing the spline to fit arbitrary function.
Trainable splines are highly flexible while maintaining computational efficiency, as the linear interpolation requires low complexity compared to other function approximators such as FC networks.
To enhance the model's adaptability to various channel qualities and transmission rates, we introduce adaptation modules in the following three aspects.

\subsection{Rate Adaptation}
\begin{figure*}[!t]
    \centering
    \includegraphics[width=1\linewidth]{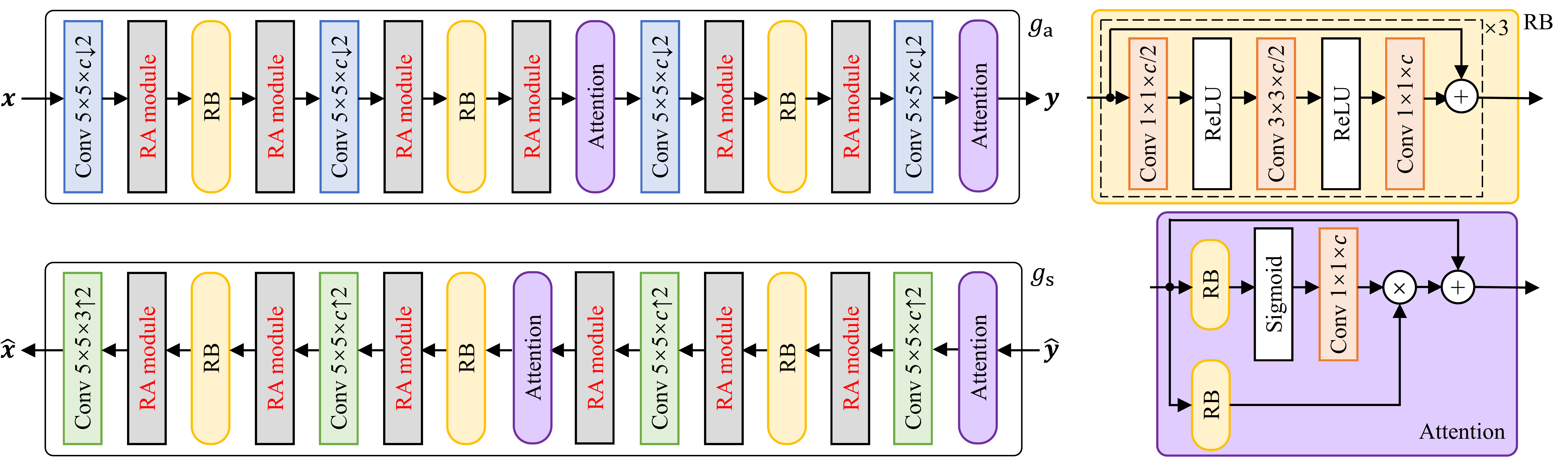}
    \caption{Network architecture of the proposed analysis and synthesis transforms (RB: Residual Bottleneck). The analysis transform $g_{\rm a}$ extracts a compact latent representation from the high-dimensional input image $\boldsymbol{x}$, and the synthesis transform $g_{\rm s}$ reconstructs the image from decoded latent features $\hat{\boldsymbol{y}}$. We implement six RA modules in both the analysis and synthesis transforms to modify hidden features to meet requirements of bandwidth. Parameters of convolutional layer are denoted as kernel height $\times$ kernel width $\times$ the number of filters, $\uparrow$ and $\downarrow$ denote up-sampling and down-sampling, respectively, followed by the stride.}
    \label{fig_NTC_model}
\end{figure*}
As illustrated in Fig. \ref{fig_NTC_model}, we implement six RA modules in both the analysis transform $g_{\rm a}$ and the synthesis transform $g_{\rm s}$ to modify hidden features to meet bandwidth requirements.
Experiments show that the entropy of latent features is approximately proportional to Lagrange multiplier $\lambda$ in a logarithmic scale.
Thus, in each RA module, channel-wise scaling factors are generated by $c$ trainable splines based on $\log \lambda$, and the hidden features in each channel are multiplied by the corresponding scaling factor as:
\begin{equation}
    \boldsymbol{y}_{\rm out} = f_{\rm RA}(\log \lambda) \odot \boldsymbol{y}_{\rm in} ,
    \label{eq_f_RA}
\end{equation}
where $\boldsymbol{y}_{\rm out}$ and $\boldsymbol{y}_{\rm in}$ are the output and the input of adaptation modules, respectively, and $f_{\rm RA}(\cdot)$ denotes the spline functions with respect to $\log \lambda$.
Each scaling factor represents the relative importance of latent features within the corresponding channel.
When bandwidth available is limited with a small Lagrange multiplier, the scaling factors assigned to features related to image details are minimized to reserve bandwidth for the overall structure.
Conversely, when bandwidth available is sufficient with a large Lagrange multiplier, the scaling factors assigned to features depicting details are increased to improve transmission quality.

\subsection{Adaptive Embedding}
\begin{figure*}[!t]
    \centering
    \includegraphics[width=.8\linewidth]{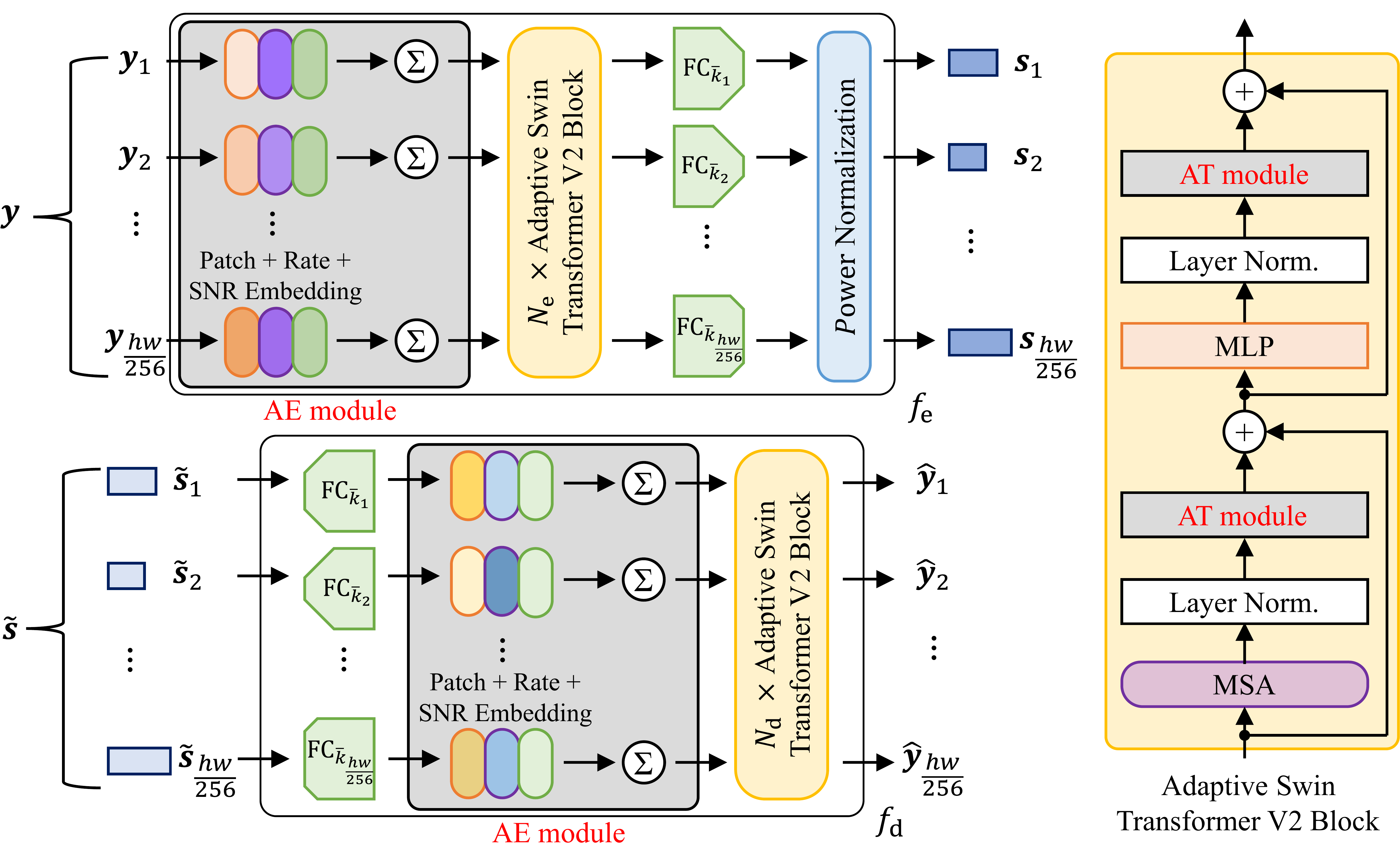}
    \caption{Network architecture of the proposed encoder and decoder networks (Layer Norm.: Layer Normalization). We employ AE modules at the beginning of both the encoder network $f_{\rm e}$ and the decoder network $f_{\rm d}$ to incorporate information regarding channel qualities and bandwidth requirements. We also introduce AT modules after each layer normalization in transformer blocks to adapt hidden features to various channel conditions and transmission rates.}
    \label{fig_codec}
\end{figure*}
As illustrated in Fig. \ref{fig_codec}, we employ AE modules at the beginning of both the encoder network $f_{\rm e}$ and the decoder network $f_{\rm d}$ to incorporate information regarding channel qualities and bandwidth requirements.
Transformer networks combine trainable embeddings with patch embeddings, \textit{i.e.}, input features after projection, to provide ensuing transformer blocks with additional information, such as spatial positions \cite{Swin}.
In NTSCC systems, the patch embeddings provided to the encoder and decoder networks contain only the semantic information of the image. 
However, other critical information, such as channel conditions and transmission rates, is not included in these patch embeddings.
To address this problem, we propose AE modules, where SNR embeddings are generated by $c$ trainable splines based on the instantaneous SNR $\gamma$ in dB scale.
Additionally, a set of trainable rate embeddings $\{ \boldsymbol{r}_{v_1}, \boldsymbol{r}_{v_2}, \ldots, \boldsymbol{r}_{v_K} \}$, each with $c$ dimensions, are implemented according to the possible symbol length set $\mathcal{K}$ \cite{NTSCC}.
The patch embeddings are combined with both the SNR embeddings and the rate embeddings as:
\begin{equation}
    \boldsymbol{y}_{i,\rm{out}} = f_{\rm AE}(\gamma) + \boldsymbol{r}_{\bar{k}_i} + \boldsymbol{y}_{i,\rm{in}} ,
    \label{eq_f_AE}
\end{equation}
where $f_{\rm AE}(\cdot)$ represents the spline functions that generate SNR embeddings, and $\boldsymbol{r}_{\bar{k}_i}$ denotes the rate embedding corresponding to the symbol length allocated to the $i$-th latent vector.

\subsection{Adaptive Transform}
As illustrated in Fig. \ref{fig_codec}, we introduce AT modules after each layer normalization in transformer blocks of the encoder and decoder networks to adapt hidden features to various channel conditions and transmission rates.
In Swin Transformer V2 blocks \cite{Swin}, layer normalization is operated after MSA and multilayer perceptron (MLP) modules.
The proposed AT module performs channel-wise affine transformation on hidden features, where both the weight vectors and the bias vectors consist of SNR components and rate components.
The SNR components are generated by $c$ trainable splines based on the instantaneous SNR $\gamma$, and the rate components are selected from a set of $c$-dimensional trainable vectors according to the allocated symbol length.
The sets of rate components for weight vectors and bias vectors are denoted as $\{ \boldsymbol{w}_{v_1}, \boldsymbol{w}_{v_2}, \ldots, \boldsymbol{w}_{v_K} \}$ and $\{ \boldsymbol{b}_{v_1}, \boldsymbol{b}_{v_2}, \ldots, \boldsymbol{b}_{v_K} \}$, respectively.
The affine transformation is given by:
\begin{equation}
    \boldsymbol{y}_{i,\rm{out}} = (f_{{\rm AT},\boldsymbol{w}}(\gamma) + \boldsymbol{w}_{\bar{k}_i}) \odot \boldsymbol{y}_{i,\rm{in}} + f_{{\rm AT},\boldsymbol{b}}(\gamma) + \boldsymbol{b}_{\bar{k}_i} ,
    \label{eq_f_AT}
\end{equation}
where $f_{{\rm AT},\boldsymbol{w}}(\cdot)$ and $f_{{\rm AT},\boldsymbol{b}}(\cdot)$ represent the spline functions that generate the SNR components of weight vectors and bias vectors, respectively, $\boldsymbol{w}_{\bar{k}_i}$ and $\boldsymbol{b}_{\bar{k}_i}$ denote the rate components of weight vectors and bias vectors corresponding to the symbol length allocated to the $i$-th latent vector, respectively.

\section{Experiment Results}
\subsection{Experiment Setup}
\subsubsection{Network Implementation}
The latent representation has a channel dimension of $c = 256$, which is partitioned into $L = 5$ groups via extensive experiments.
Each group contains $c_l = 48$ $(l = 1, 2, 3, 4)$ channels, except for the last group, which has $c_5 = 64$ channels.
The possible symbol lengths are set as $\mathcal{K} = \{ 8, 16, 24, 32, 48, 64, 80, 96, 112, 128, 144, 160, 184, 208, 232, \\ 256 \}$.
The splines in RA modules have $N_{\rm RA} = 15$ knots, with the input $\log_{10} \lambda$ uniformly distributed between $-2$ and $0$.
The splines in AE modules and AT modules both have $N_{\rm AE} = N_{\rm AT} = 11$ knots, with the input $\gamma$ uniformly distributed between $0$ and $20$.
The encoder and decoder networks both employ $N_{\rm e} = N_{\rm d} = 4$ Swin Transformer V2 blocks \cite{Swin} which have a window size of $8 \times 8$, 
with the dimension of each patch embedding the same as that of the latent representation $c = 256$.
All neural networks are implemented using the PyTorch library \cite{Pytorch}.

\subsubsection{Training Details}
The dataset for training consists of 41,620 images with a minimum resolution of $1024 \times 512$ pixels from the validation set of Open Images \cite{OpenImages}.
During model training, images are randomly cropped into patches of $512 \times 512$ pixels.
We employ the Adam optimizer \cite{Adam} with a learning rate of $10^{-4}$ and a batch size of 16.
In accordance with previous NTC models \cite{Minnen2020, ELIC}, the learnable quantiles of probability model are optimized with a learning rate ten times that of other parameters.
In each training iteration, SNR $\gamma$ is randomly sampled from 0 dB to 20 dB, bandwidth coefficient $\eta$ is randomly sampled from 0.1 to 0.3, and Lagrange multiplier $\lambda$ is randomly sampled in logarithmic scale from 0.01 to 1.
For the proposed MR-NTSCC model with no adaptation modules, SNR is fixed at 10 dB, and Lagrange multiplier maintains at 0.1 during training.

The proposed MA-NTSCC model is trained in two stages.
In the first stage, the NTC model is trained individually by directly feeding the latent features $\boldsymbol{y}$ into the synthesis transform $g_{\rm s}$, which bypasses channel impairment as well as the encoder and decoder networks.
The loss function is expressed as
\begin{equation}
    \mathcal{L}_{\rm NTC} = -\eta \sum_{i} \log p_{\boldsymbol{y}_i | \boldsymbol{z}}(\boldsymbol{y}_i | \boldsymbol{z}) + \lambda d(\boldsymbol{x}, \hat{\boldsymbol{x}}_{\rm NTC}) ,
    \label{eq_loss_NTC}
\end{equation}
where $\hat{\boldsymbol{x}}_{\rm NTC}$ denotes the image decoded directly by the synthesis transform.
We choose mean square error (MSE) as the distortion function to optimize PSNR performance.
The NTC model is trained via $120$ epochs, and the learning rate is divided by a factor of $\sqrt{10}$ every 10 epochs during the last 40 epochs. 
In the second stage, the proposed MA-NTSCC model is trained end-to-end via $280$ epochs to achieve convergence, with the learning rate similarly divided by a factor of $\sqrt{10}$ every 10 epochs during the last 40 epochs.
Following NTSCC \cite{NTSCC}, the loss function is a combination of NTSCC and NTC losses, given in (\ref{eq_loss_NTSCC}) and (\ref{eq_loss_NTC}), respectively, to stabilize training and maintain the performance of the NTC model, expressed as
\begin{equation}
    \mathcal{L} = \mathcal{L_{\rm NTSCC}} + \theta \mathcal{L_{\rm NTC}} ,
    \label{eq_loss}
\end{equation}
where $\theta$ is the weighting coefficient of the NTC loss.
In the first 200 epochs, the coefficient is set as $\theta = 1$.
From epochs 201 to 240, the weighting coefficient is multiplied by a factor of 0.944 in each epoch, and is maintained at $\theta = 0.1$ afterwards. 

\subsubsection{Evaluation Details}
To demonstrate the adaptability of the proposed MA-NTSCC model across diverse datasets, we evaluate the wireless image transmission performance on three datasets: Kodak \cite{Kodak}, the training set of CLIC2020 \cite{CLIC} and the validation set of Cityscape \cite{Cityscape}.
Kodak dataset consists of 24 photographs, depicting distinct scenes with a medium resolution of $768 \times 512$ pixels.
The training set of CLIC2020 contains 585 images with varying styles and resolutions ranging from $600 \times 400$ to $2048 \times 1945$ pixels.
The validation set of Cityscape comprises 500 street-view photographs with a high resolution of $2048 \times 1024$ pixels.
Note that we employ identical weights in the proposed MA-NTSCC and MR-NTSCC models during the entire evaluation process, instead of training separate models with different hyperparameters to enhance performance.
To minimize performance fluctuation caused by random channel realizations, each image is transmitted 10 times, and the averaged values are reported as the final results.

\subsection{Results Analysis}
\subsubsection{RD Performance}
\begin{figure*} [!t]
    \centering
    \subfloat[]{\includegraphics[width=.333\linewidth]{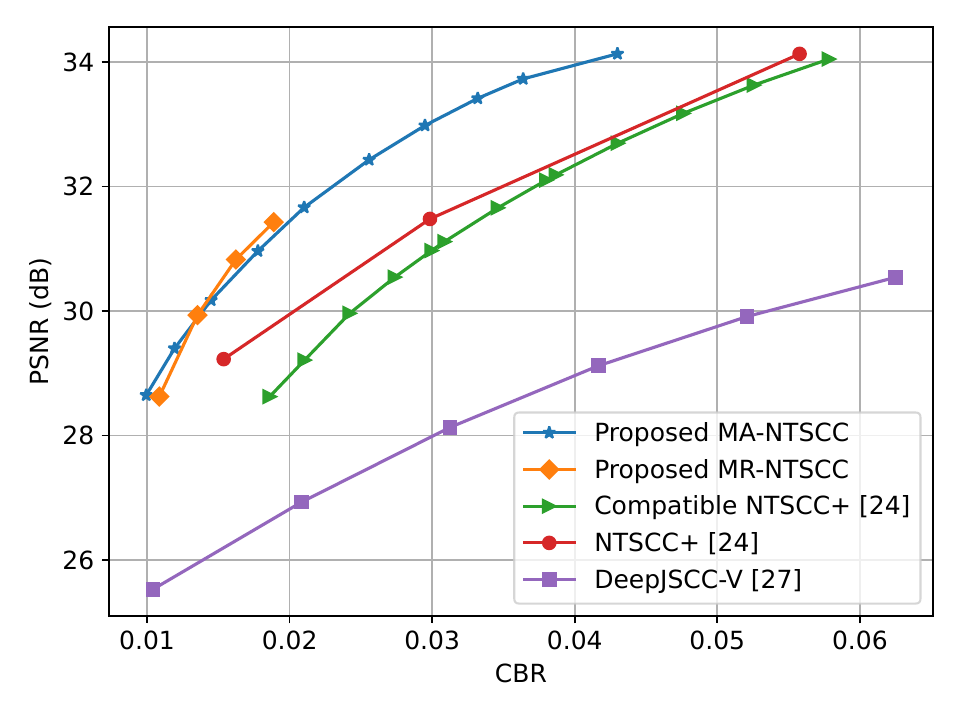}}
    \subfloat[]{\includegraphics[width=.333\linewidth]{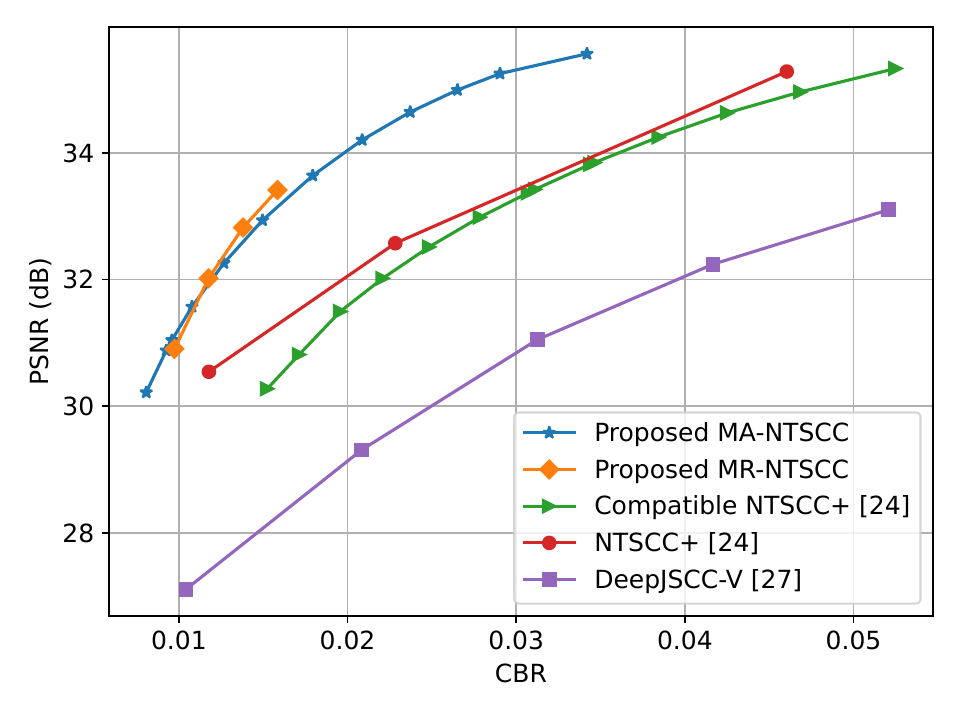}}
    \subfloat[]{\includegraphics[width=.333\linewidth]{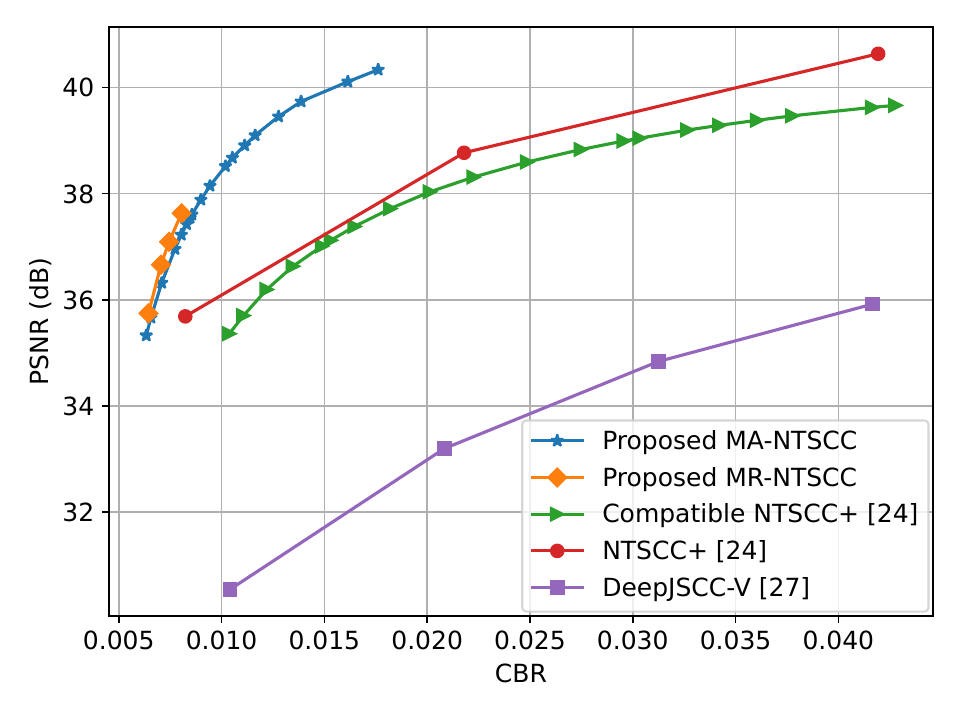}}
    \caption{Impact of CBR on PSNR performance of the proposed MA-NTSCC and MR-NTSCC models in AWGN channel with SNR $\gamma = 10 \; \rm{dB}$. (a), (b) and (c) show results on Kodak, CLIC2020 and Cityscape datasets, respectively.}
    \label{fig_RD_PSNR}
\end{figure*}
\begin{figure*} [!t]
    \centering
    \subfloat[]{\includegraphics[width=.333\linewidth]{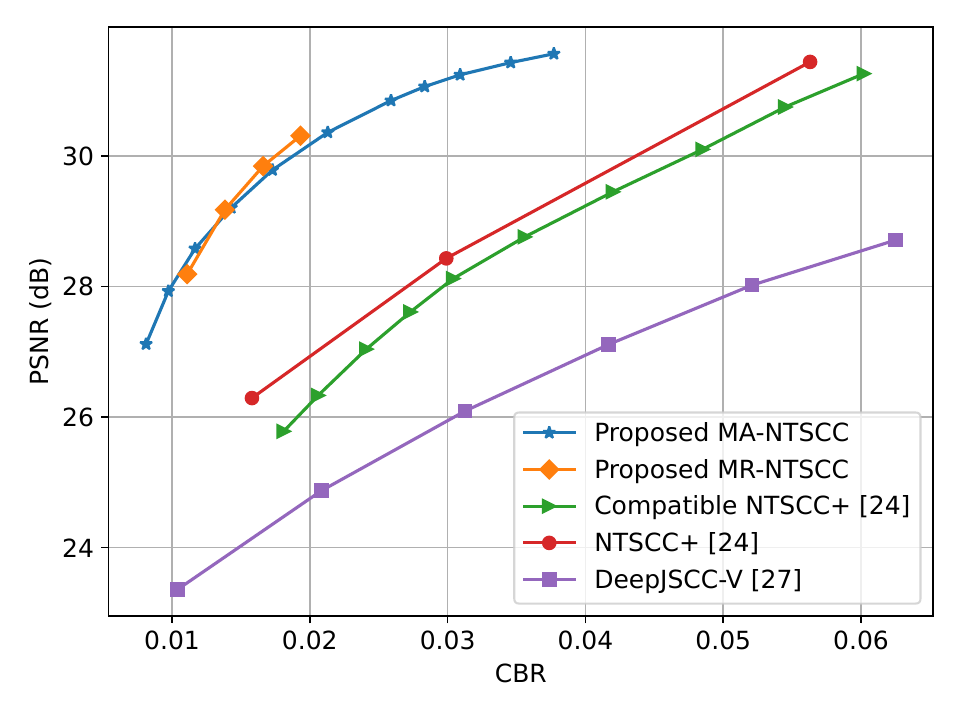}}
    \subfloat[]{\includegraphics[width=.333\linewidth]{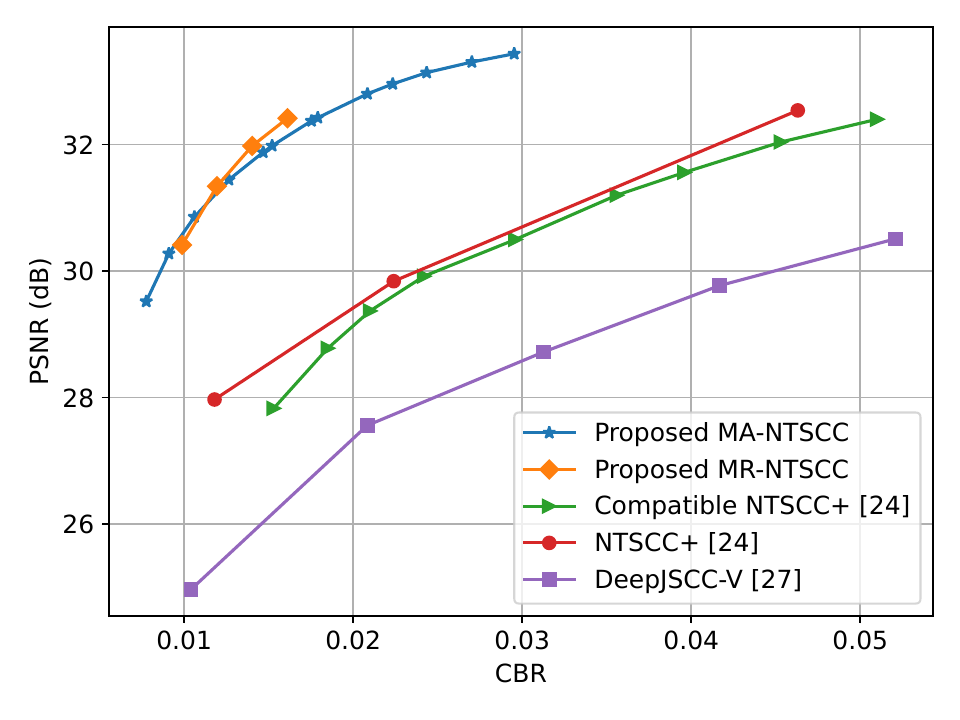}}
    \subfloat[]{\includegraphics[width=.333\linewidth]{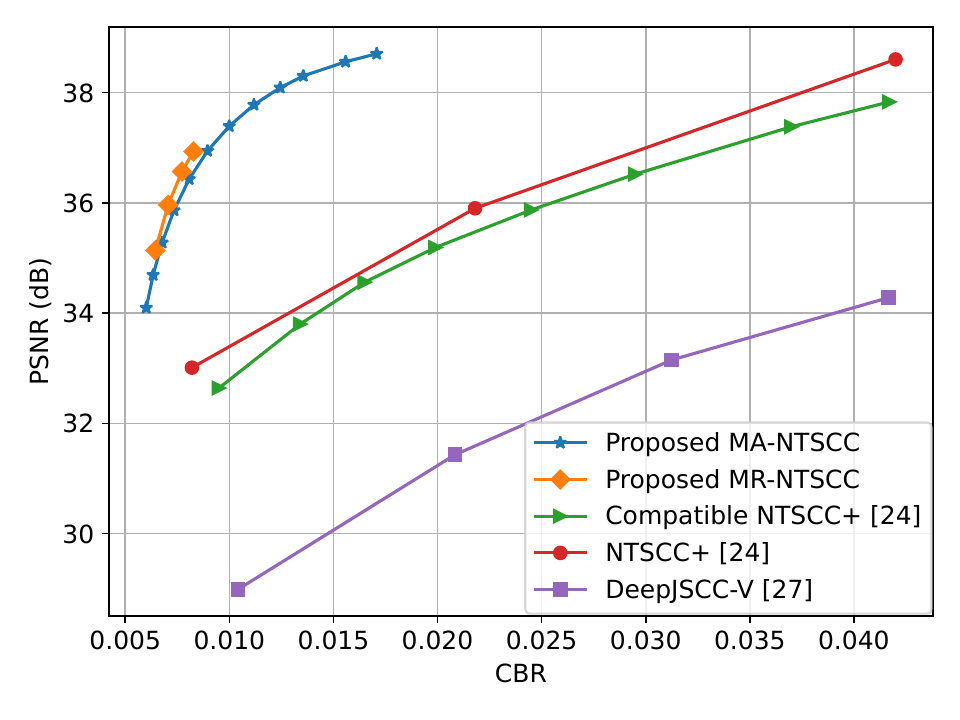}}
    \caption{Impact of CBR on PSNR performance of the proposed MA-NTSCC and MR-NTSCC models in Rayleigh fading channel with SNR $\gamma = 10 \; \rm{dB}$. (a), (b) and (c) show results on Kodak, CLIC2020 and Cityscape datasets, respectively.}
    \label{fig_RD_Rayleigh}
\end{figure*}
Figs. \ref{fig_RD_PSNR} and \ref{fig_RD_Rayleigh} show the impact of CBR on PSNR performance of the proposed MA-NTSCC and MR-NTSCC models in AWGN channel and Rayleigh fading channel, respectively, with SNR $\gamma = 10 \; \rm{dB}$.
To traverse the RD curve, the bandwidth coefficient $\eta$ of both the proposed MA-NTSCC and MR-NTSCC models varies from 0.1 to 0.4.
Additionally, a Lagrange multiplier $\lambda$, ranging from 0.01 to 1, is provided to the RA modules of the proposed MA-NTSCC model.
Results show that the proposed MA-NTSCC model consumes significantly less bandwidth while maintaining equivalent transmission quality for all three datasets, compared with the compatible NTSCC+ model \cite{NTSCC+}.
For medium-resolution images in Kodak dataset, the proposed MA-NTSCC model requires only around 60\% of the bandwidth for a target PSNR from 28.6 dB to 33 dB in AWGN channel, or less than half of the bandwidth for a target PSNR from 27.2 dB to 31.2 dB in Rayleigh fading channel, in comparison to the compatible NTSCC+ model \cite{NTSCC+}.
For high-resolution images in Cityscape dataset, the proposed MA-NTSCC model achieves a higher reduction in bandwidth, requiring less than half of the bandwidth for a target PSNR exceeding 37.5 dB in AWGN channel, or less than 40\% of the bandwidth for a target PSNR exceeding 34 dB in Rayleigh fading channel, compared with the compatible NTSCC+ model \cite{NTSCC+}.
These bandwidth savings verify the effectiveness of the proposed multi-reference entropy model, which leverages correlations in images to allocate bandwidth more efficiently, compared with the existing models \cite{NTSCC+}.
The DeepJSCC-V model \cite{DJSCC-V} delivers performance inferior to NTSCC-based models, as uniform rate is allocated for all latent vectors, disregarding the significant variation in semantic complexity between different regions of the image.
Moreover, the achievable CBRs of the DeepJSCC-V model are discrete values with a interval of $1/96$, less adaptive than the proposed MA-NTSCC model and the compatible NTSCC+ model \cite{NTSCC+} with continuous achievable CBRs.

Furthermore, the proposed RA module enables a single model to deliver competitive performance across a wide range of bandwidth ratios.
As shown in Figs. \ref{fig_RD_PSNR} and \ref{fig_RD_Rayleigh}, the proposed MA-NTSCC model accommodates a broader range of bandwidth ratios while maintaining similar performance, compared with the proposed MR-NTSCC model trained under a specific Lagrange multiplier $\lambda$ that determines the trade-off between rate and distortion.
Conversely, the compatible NTSCC+ model \cite{NTSCC+} achieves PSNR performance inferior to the separately trained NTSCC+ models \cite{NTSCC+}, especially at both lower and upper CBR ranges.

\subsubsection{SNR-Distortion Performance}
\begin{figure*} [!t]
    \centering
    \subfloat[]{\includegraphics[width=.333\linewidth]{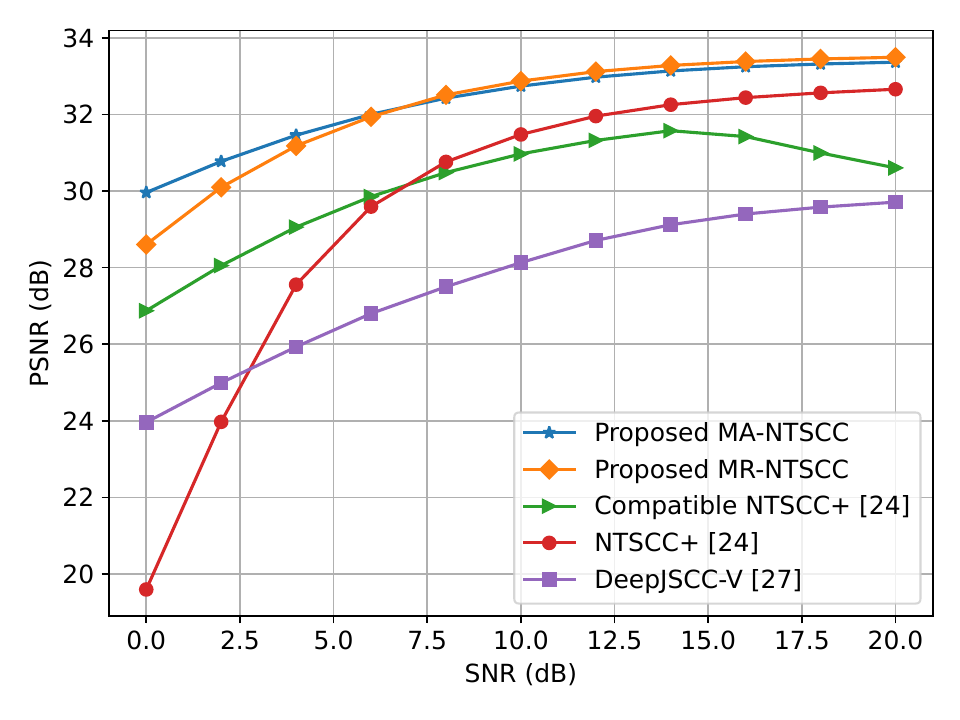}}
    \subfloat[]{\includegraphics[width=.333\linewidth]{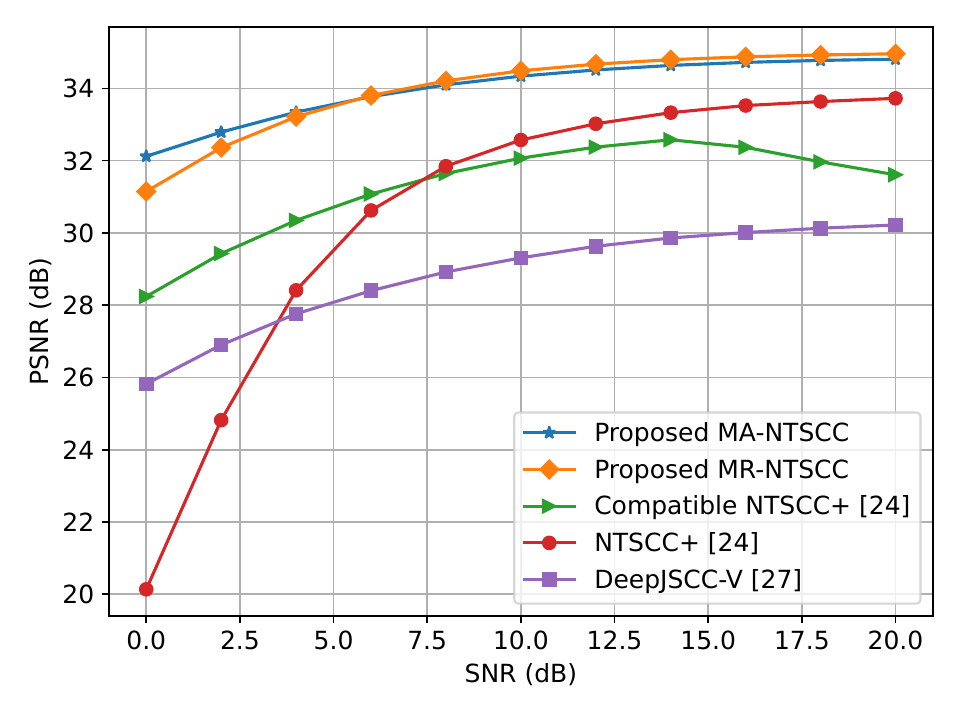}}
    \subfloat[]{\includegraphics[width=.333\linewidth]{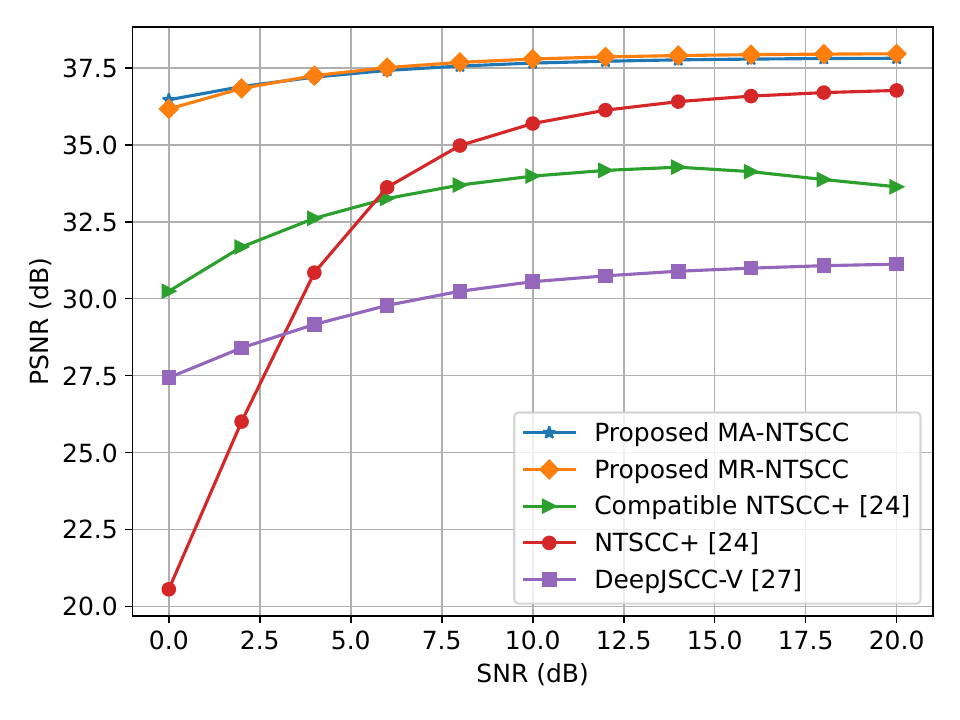}}
    \caption{Impact of SNR on PSNR performance of the proposed MA-NTSCC and MR-NTSCC models in AWGN channel under identical CBR. (a), (b) and (c) show the results on Kodak, CLIC2020 and Cityscape datasets, respectively.}
    \label{fig_SD_PSNR}
\end{figure*}
\begin{figure*} [!t]
    \centering
    \subfloat[]{\includegraphics[width=.333\linewidth]{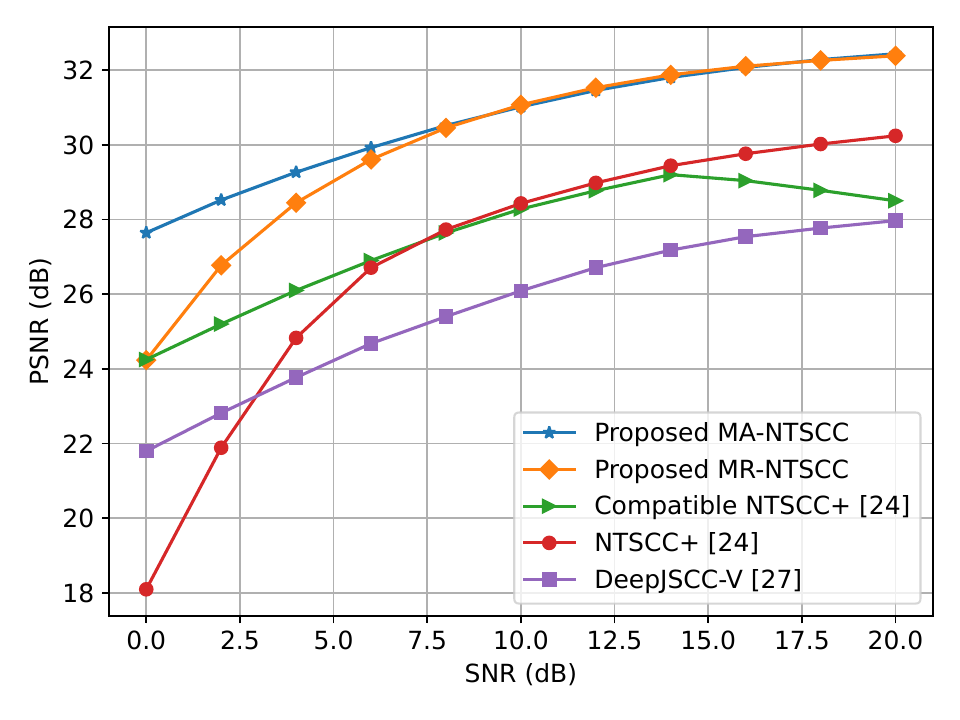}}
    \subfloat[]{\includegraphics[width=.333\linewidth]{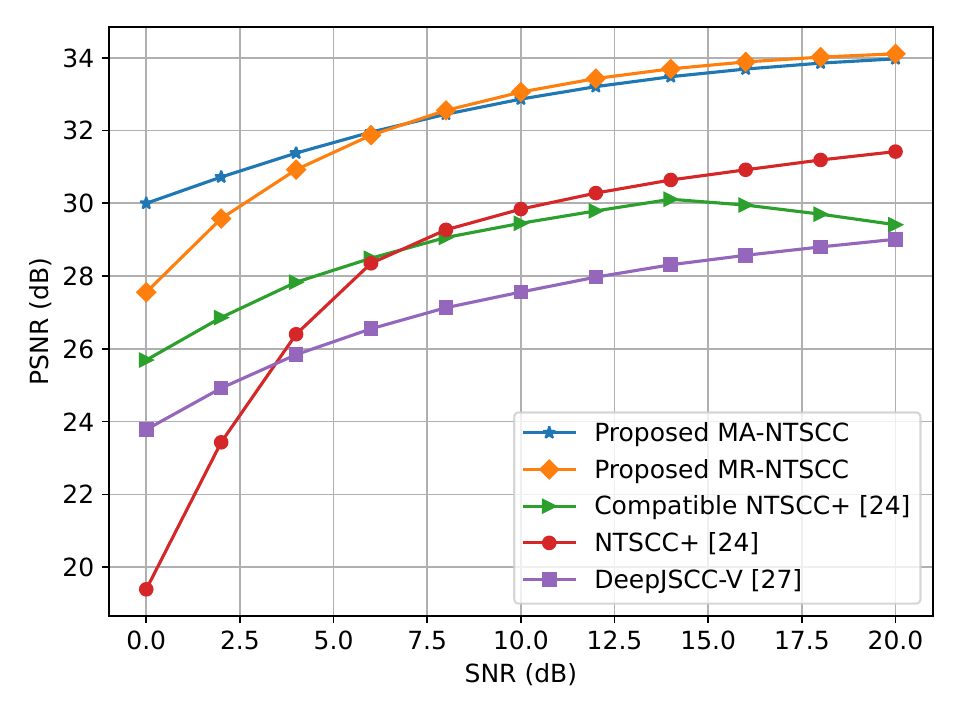}}
    \subfloat[]{\includegraphics[width=.333\linewidth]{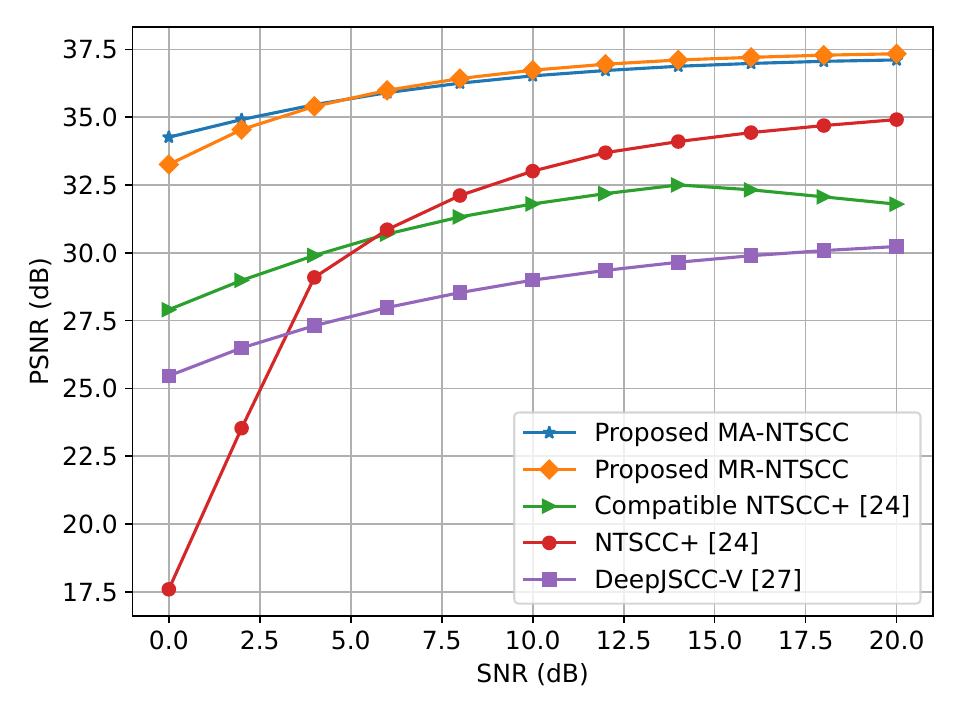}}
    \caption{Impact of SNR on PSNR performance of the proposed MA-NTSCC and MR-NTSCC models in Rayleigh fading channel under identical CBR. (a), (b) and (c) show the results on Kodak, CLIC2020 and Cityscape datasets, respectively.}
    \label{fig_SD_Rayleigh}
\end{figure*}
Figs. \ref{fig_SD_PSNR} and \ref{fig_SD_Rayleigh} show the impact of SNR on PSNR performance of the proposed MA-NTSCC and MR-NTSCC models in AWGN channel and Rayleigh fading channel, respectively, under identical CBR.
We carefully select bandwidth coefficient $\eta$ and Lagrange multiplier $\lambda$ to ensure identical CBR, with the NTSCC+ model trained with a fixed Lagrange multiplier of 0.0483.
Results show that the proposed MA-NTSCC model achieves significant improvements in transmission quality under identical CBR across various SNRs and all three datasets, compared with the compatible NTSCC+ model \cite{NTSCC+}.
For medium-resolution images in Kodak dataset, the proposed MA-NTSCC model enhances PSNR performance by 3 dB in low and high SNR ranges, and by 2 dB in medium SNR regimes, in AWGN channel and compared with the compatible NTSCC+ model \cite{NTSCC+}.
For high-resolution images in Cityscape dataset, the proposed MA-NTSCC model achieves greater PSNR improvements of 6 dB in low SNR regimes and 4 dB in medium and high SNR ranges, in AWGN channel, compared with the compatible NTSCC+ model \cite{NTSCC+}.
The DeepJSCC-V model \cite{DJSCC-V} delivers performance inferior to NTSCC-based models, as uniform rate is allocated for all latent vectors, disregarding the significant variation in semantic complexity between different regions of the image.
In Rayleigh fading channel, the PSNR improvements of the proposed MA-NTSCC model are even more prominent than those of the NTSCC+ model \cite{NTSCC+} and the DeepJSCC-V model \cite{DJSCC-V}, ranging from 2.2 dB to 3.9 dB for Kodak dataset, and from 4.4 dB to 6.4 dB for Cityscape dataset.
The superior transmission quality verifies the adaptability of the proposed encoder and decoder networks across various channel conditions.

The transmission quality of traditional NTSCC models, trained under a relatively high SNR, significantly degrades in a low SNR regime, as shown in the performance curves of the NTSCC+ model in Figs. \ref{fig_SD_PSNR} and \ref{fig_SD_Rayleigh}.
To overcome this limitation, the proposed AE and AT modules enable the proposed MA-NTSCC model to adapt hidden features to dynamic SNRs.
Consequently, the proposed MA-NTSCC model enhances transmission quality in low SNR regimes, without compromising performance in medium and high SNR regimes, compared with the proposed MR-NTSCC model trained under a fixed SNR of 10 dB.
Conversely, the compatible NTSCC+ model \cite{NTSCC+} improves transmission quality in low SNR regimes at the expense of the degraded performance in medium and high SNR regimes, in comparison to the NTSCC+ model \cite{NTSCC+} trained under a fixed SNR of 10 dB.
Moreover, the proposed MA-NTSCC model can accommodate a wide range of SNRs from 0 dB to 20 dB, while the compatible NTSCC+ model \cite{NTSCC+} can only adapt to a narrow range of SNRs from 0 dB to 14 dB, resulting in performance degradation when SNR exceeds 14 dB.

\subsubsection{Visual Comparison}
\begin{figure*}[!t]
    \centering
    \includegraphics[width=1\linewidth]{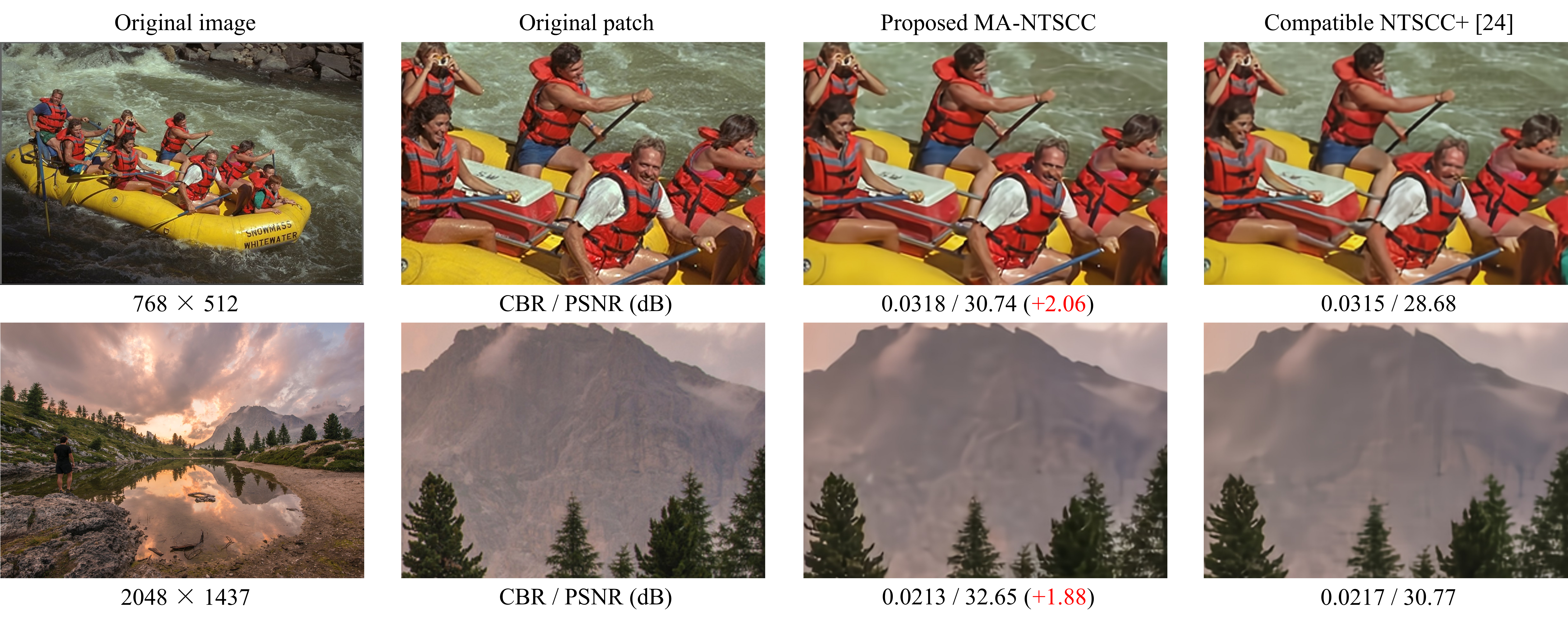}
    \caption{Examples of the reconstructed images from Kodak and CLIC2020 datasets, respectively. The first column shows the original images. The second column shows the cropped patches in original images. The third and fourth columns show the cropped patches in the reconstructed images from the proposed MA-NTSCC model and the compatible NTSCC+ model \cite{NTSCC+}, respectively, under similar CBRs over AWGN channel with SNR $\gamma = 10 \; \rm{dB}$.}
    \label{fig_samples}
\end{figure*}
Fig. \ref{fig_samples} presents examples of the reconstructed images from the proposed MA-NTSCC model and the compatible NTSCC+ model \cite{NTSCC+} under similar CBRs over AWGN channel with SNR $\gamma = 10 \; \rm{dB}$, respectively. 
The compatible NTSCC+ model \cite{NTSCC+} is used as a comparison to show the improvements in transmission quality of the proposed MA-NTSCC model as below. 
In the first row of examples, the image reconstructed by the proposed MA-NTSCC model exhibits finer details on the facial expressions, significantly improving perceptual quality.
The second row of examples demonstrate that the proposed MA-NTSCC model better preserves the texture of mountains in the background and the shape of leaves in the foreground.
These perceptual observations are consistent with the improvements in objective PSNR performance.

\subsubsection{Trained Spline Functions}
\begin{figure} [!t]
    \centering
    \subfloat[]{\includegraphics[width=.8\linewidth]{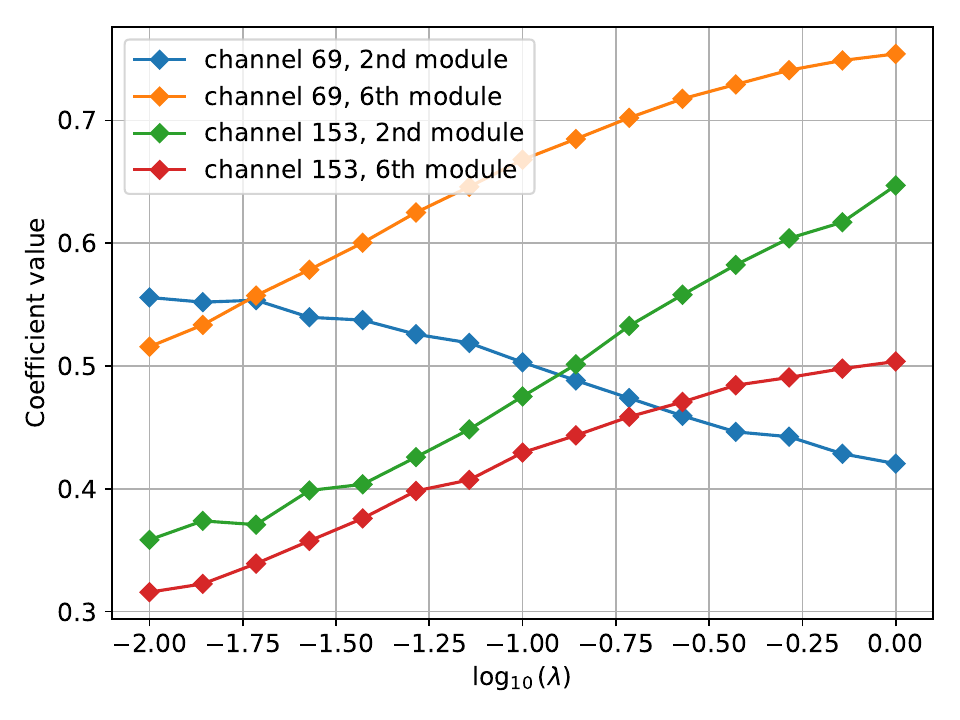}} \hfill
    \subfloat[]{\includegraphics[width=.8\linewidth]{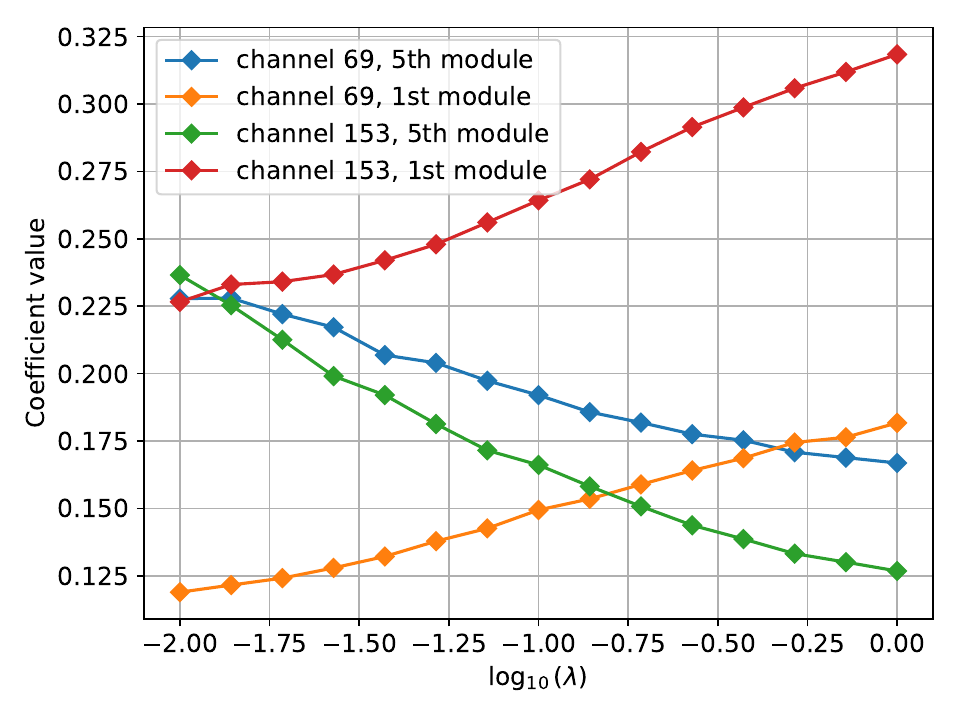}}
    \caption{Trained spline functions of the proposed RA modules in the NTC model. (a) and (b) illustrate functions of the proposed RA modules in the analysis and synthesis transforms, respectively.}
    \label{fig_RA}
\end{figure}
Fig. \ref{fig_RA} illustrates the trained spline functions of the proposed RA modules in the NTC model.
We randomly select indexes of the channel dimension and the RA module in the analysis transform $g_{\rm a}$. 
The selected channels of the corresponding RA modules in the synthesis transform $g_{\rm s}$ are also demonstrated.
The proposed RA module assigns distinct weights to features in each channel based on Lagrange multiplier $\lambda$ which determines the RD trade-off.
The channel-wise weights in an RA module notably vary with respect to both Lagrange multiplier and channel index, thereby enabling a single model to perform competitively across various bandwidth ratios.
The trained spline functions in the proposed RA modules are generally monotonic, representing changes in the relative importance of the features within the channel, 
as discussed in Subsection IV-A.
Additionally, the weights generated by the RA module in the analysis transform are not correlated with those generated by the corresponding RA module in the synthesis transform, due to the characteristics of CNNs, where the calculations of output features across different channels are homogeneous.
Consequently, the hidden features within a specific channel generated by different layers of CNN do not exhibit greater relevance compared to features from different channels.

\begin{figure} [!t]
    \centering
    \subfloat[]{\includegraphics[width=.8\linewidth]{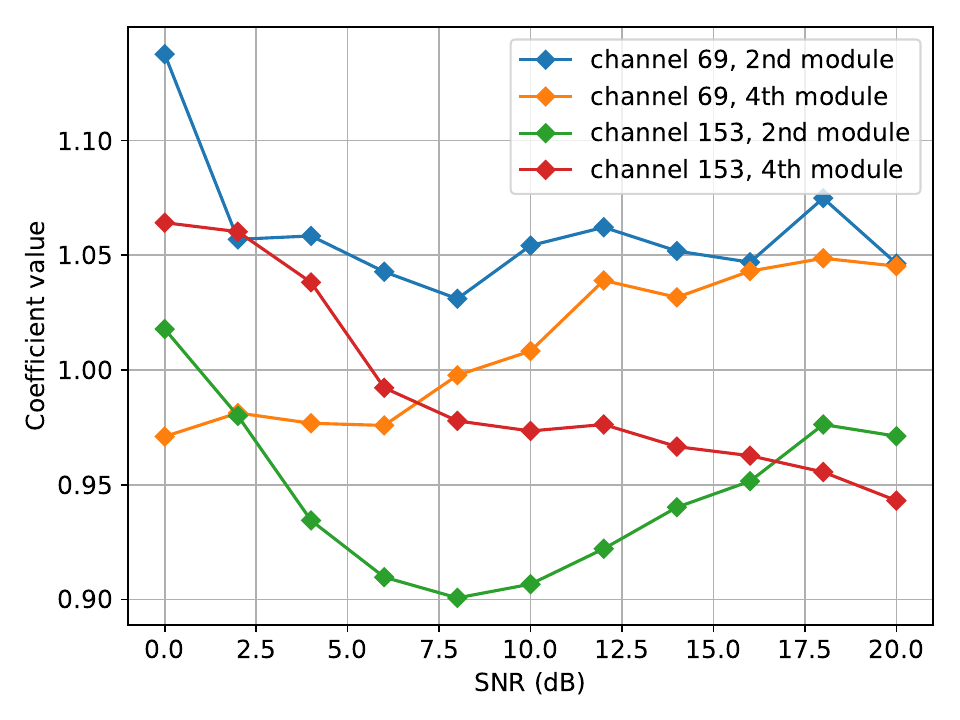}} \hfill
    \subfloat[]{\includegraphics[width=.8\linewidth]{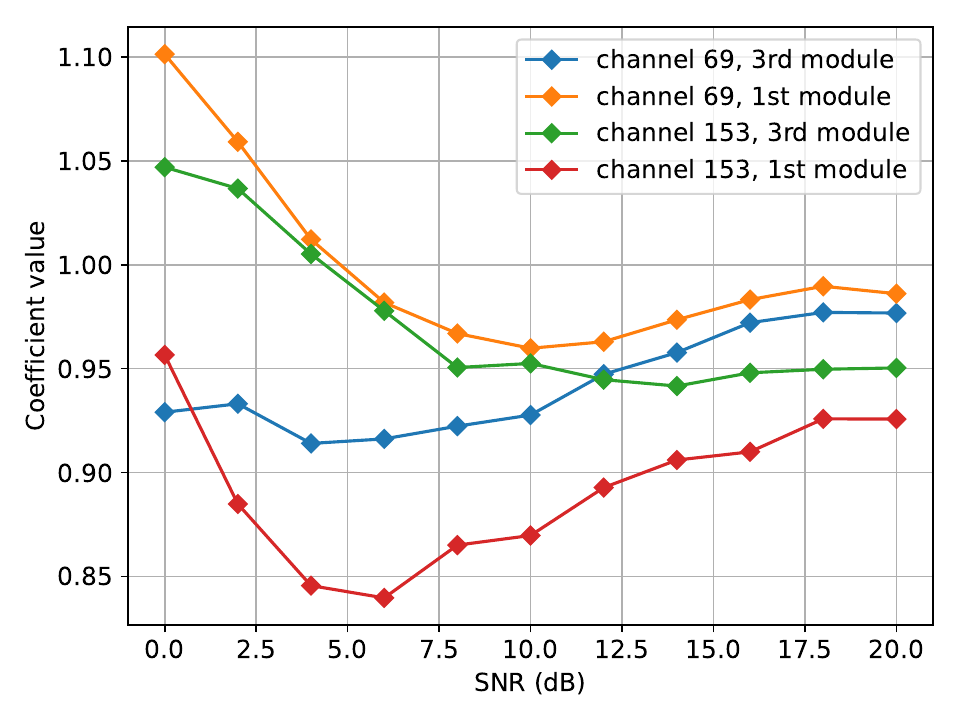}}
    \caption{Trained spline functions of the proposed AT modules in the encoder and decoder networks. (a) and (b) illustrate functions of the proposed AT modules in the encoder and decoder networks, respectively.}
    \label{fig_AT}
\end{figure}
Fig. \ref{fig_AT} illustrates the trained spline functions of the proposed AT modules in the encoder and decoder networks.
We present the weights generated by the spline function based on SNR in the proposed AT module following the MSA module.
Results show that the proposed AT module modifies the hidden features to enhance robustness against channel impairments, thereby maintaining transmission quality across various SNRs.
The weights of the proposed AT module are close to unity, exhibiting smaller variations compared with those of the proposed RA module.
The reduced discrepancy in the weights aligns with the observation that hidden features generated by models trained under different SNRs demonstrate higher cosine similarity than those generated by models trained under varying Lagrange multipliers \cite{NTSCC+}.
Additionally, the weights generated by the AT module in the encoder network are not correlated with those generated by the corresponding AT module in the decoder network, as the architectures of the encoder and decoder networks are not symmetric.

\subsection{Ablation Study}
\begin{figure} [!t]
    \centering
    \includegraphics[width=.9\linewidth]{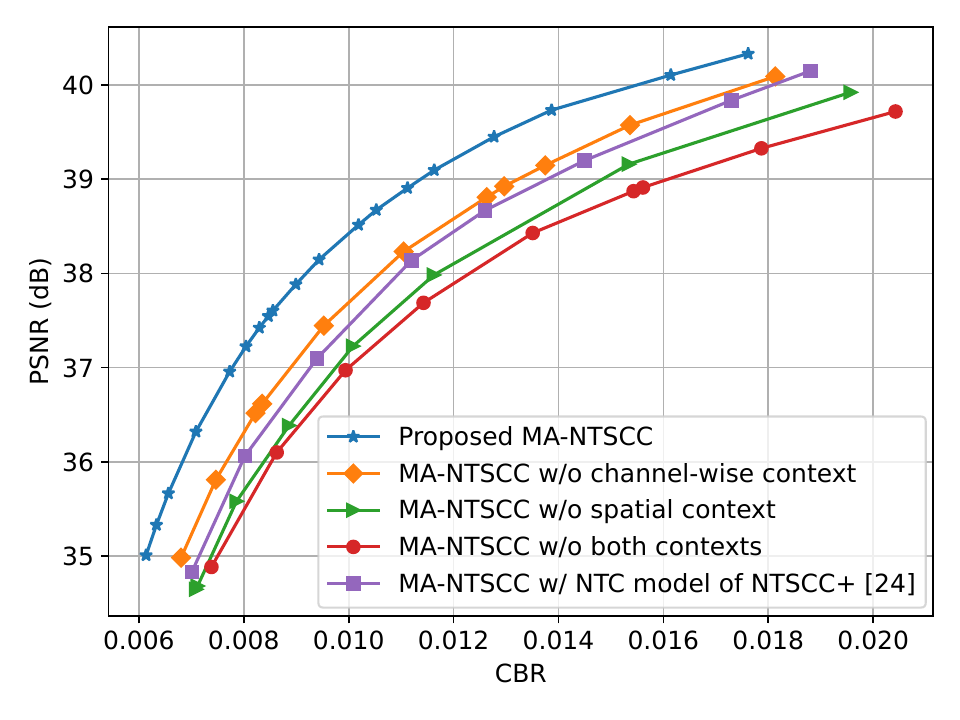}
    \caption{Impact of CBR on PSNR performance of the proposed MA-NTSCC models with alternative NTC models on Cityscape dataset (w: with, w/o: without).}
    \label{fig_ablation_ent}
\end{figure}
To verify the effect of the proposed multi-reference entropy model on conserving bandwidth, we conduct ablation study on the spatial and channel-wise contexts of the proposed MA-NTSCC model. 
Specifically, we remove the proposed subnets $g_{\rm sp}$ and $g_{\rm ch}$ which analyze the semantic information of correlated features in spatial and channel dimensions, respectively, and reduce the dimension of the input features of the subnet $g_{\rm ep}$ accordingly.
We also replace our multi-reference model by the NTC model with checkerboard context employed in the NTSCC+ model \cite{NTSCC+}.

Fig. \ref{fig_ablation_ent} demonstrates the impact of CBR on PSNR performance of the proposed MA-NTSCC models without spatial and/or channel-wise contexts.
Results show that both the proposed spatial and channel-wise contexts contribute to bandwidth savings across various target PSNRs.
The proposed spatial context alone achieves approximately 
16\% reduction in bandwidth requirements.
The spatial context achieves high bandwidth savings to transmit high-resolution images by exploiting the mutual information between neighboring patches of images.
Similarly, the proposed channel-wise context achieves 
approximately 10\% reduction in bandwidth requirements.
Using spatial and channel-wise contexts, the proposed multi-reference entropy model reduces bandwidth consumption by approximately 25\% across various PSNRs for both datasets.
The NTC model employed in the NTSCC+ model \cite{NTSCC+} degrades performance compared to the proposed multi-reference model, validating the superiority of our design.

\begin{figure} [!t]
    \centering
    \includegraphics[width=.8\linewidth]{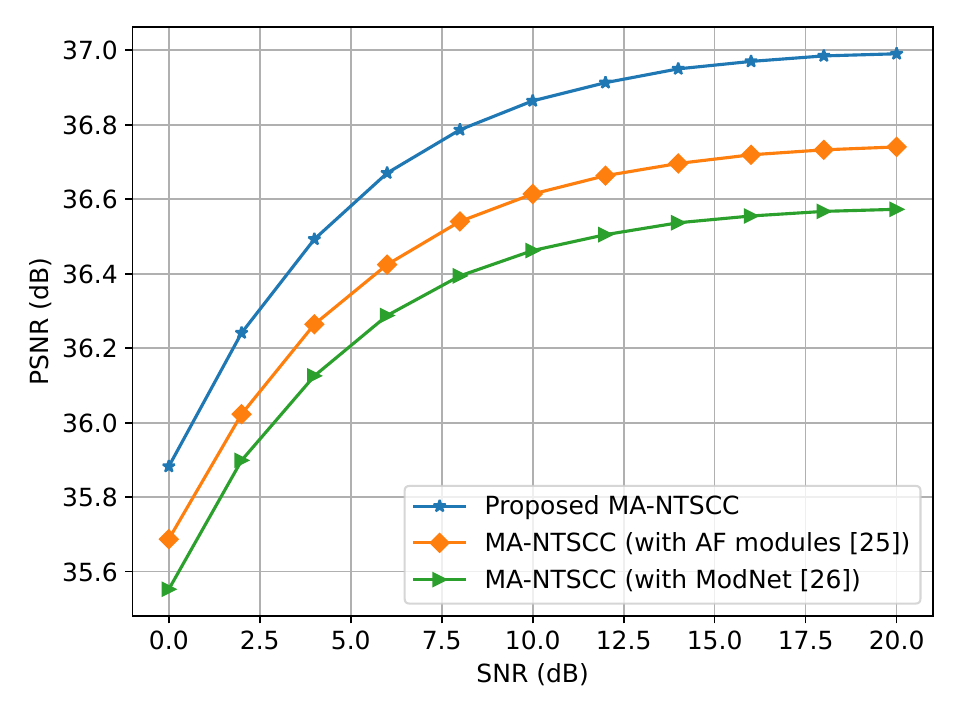}
    \caption{Impact of SNR on PSNR performance of the proposed MA-NTSCC models with alternative SNR-adaptation modules on Cityscape dataset.}
    \label{fig_ablation_ada}
\end{figure}
To verify the effect of the proposed AE and AT modules on adapting to various SNRs, we conduct ablation study on the SNR-adaptation modules.
Specifically, we remove the SNR embeddings $f_{\rm AE}(\gamma)$ in (\ref{eq_f_AE}) from the proposed AE modules, and remove the AT modules in the encoder and decoder networks of the proposed MA-NTSCC model.
Subsequently, for the MA-NTSCC model with AF modules \cite{ADJSCC}, eight AF modules are introduced into the encoder and decoder networks, and are positioned identically to the proposed AT modules.
For the MA-NTSCC model with ModNet \cite{ModNet}, we follow the original implementation, inserting ModNet after the transformer blocks in the encoder network, and before the transformer blocks in the decoder network.
Fig. \ref{fig_ablation_ada} illustrates the impact of SNR on PSNR performance of the proposed MA-NTSCC models with alternative SNR-adaptation modules on Cityscape dataset.
Using the identical proposed MA-NTSCC model, the proposed AT modules achieves transmission qualities across all SNRs superior to AF modules \cite{ADJSCC} and ModNet \cite{ModNet}. 
The PSNR performance gap between the proposed AT modules and AF modules \cite{ADJSCC} ranges from 0.2 dB in low SNR regimes to 0.3 dB in high SNR regimes. 
ModNet performs worst, with the PSNR performance gap between the proposed AT modules and ModNet \cite{ModNet} ranging from 0.3 dB in low SNR regimes to 0.4 dB in high SNR regimes.
The proposed AT module is based on trainable splines, which requires a smaller number of parameters and reduces computational complexity by more than an order of magnitude, compared with the FC networks of AF modules \cite{ADJSCC} and ModNet \cite{ModNet}.

\subsection{Complexity Analysis}
\begin{table}[!t]
    \centering
    \caption{Complexity Analysis\label{tab_complexity}}
    \begin{tabular}{|c|c|c|c|}
        \hline
        \textbf{Models} & \textbf{Model size} & \textbf{Runtime} & \textbf{Training time}\\
        \hline
        Proposed MR-NTSCC & 187.0 MB & 605.7 s & 1612.7 s \\
        \hline
        Proposed MA-NTSCC & 188.1 MB & 611.6 s & 1633.9 s \\
        \hline
        NTSCC+ \cite{NTSCC+} & 257.7 MB & 901.6 s & 2137.8 s \\
        \hline
        Compatible NTSCC+ \cite{NTSCC+} & 269.4 MB & 927.9 s & 2209.4 s \\
        \hline
    \end{tabular}
\end{table}
Table \ref{tab_complexity} demonstrates the complexity of the proposed MA-NTSCC model in terms of model size, runtime and training time, in comparison to the NTSCC+ model \cite{NTSCC+}.
The runtime is assessed by evaluating SNR-distortion performance on CLIC2020 dataset. The training time represents the average time of one training epoch.
All experiments are conducted on a server equipped with an Intel Gold 6258R CPU and four NVIDIA RTX 4090 GPUs to ensure a fair comparison.
The results indicate that the proposed MA-NTSCC model delivers superior RD performance with a 30\% reduction in model size and a 34\% decrease in runtime, compared with the compatible NTSCC+ model \cite{NTSCC+}.
These reductions highlight the efficiency of the proposed network design, and validate that the proposed multi-dimensional contexts introduce marginal overhead.
Additionally, the proposed adaptation modules in the proposed MA-NTSCC model achieves adaptability across various SNRs and bandwidth ratios with a marginal increase in model size and runtime.
The proposed adaptation modules based on trainable splines are highly efficient, with only 9.4\% of the model size and 22\% of the runtime compared with the modules introduced in the compatible NTSCC+ model \cite{NTSCC+}.

\section{Conclusions}
In this paper, we have proposed an MA-NTSCC system to improve RD performance by introducing multi-dimensional contexts into the entropy model of the NTSCC system \cite{NTSCC}. 
The proposed multi-reference entropy model leverages correlations within the latent representation in both spatial and channel dimensions.
Taking mutual information into account, the entropy model provides an accurate estimation on the entropy, which enables efficient bandwidth allocation and enhances RD performance. 
Additionally, the proposed lightweight adaptation modules enable the proposed MA-NTSCC model to achieve transmission quality comparable to separately trained models across various channel conditions and bandwidth requirements.
The proposed adaptation modules leverage trainable splines, which offer sufficient expressive power while maintaining low complexity.
Experiment results show that the proposed MA-NTSCC model achieves up to 50\% bandwidth savings at identical transmission quality, or up to 6 dB improvement in PSNR with equal SNR and bandwidth consumption, compared with the compatible NTSCC+ model \cite{NTSCC+}.
Complexity analysis of the proposed MR-NTSCC model shows a 30\% reduction in model size and a 34\% decrease in runtime, compared with the compatible NTSCC+ model \cite{NTSCC+}.

\bibliographystyle{IEEEtran}






\vfill

\end{document}